\documentclass[prl,twocolumn,superscriptaddress]{revtex4}
\usepackage{epsfig}
\usepackage{times}
\usepackage{amsmath}
\usepackage{amsfonts}
\usepackage{amssymb}
\usepackage[usenames]{color}
\usepackage{graphicx}
\newcommand{\beq}{\begin{equation}}
\newcommand{\eeq}{\end{equation}}
\newcommand{\beqa}{\begin{eqnarray}}
\newcommand{\eeqa}{\end{eqnarray}}

\newcommand{\red}{\textcolor{black}}

\newcommand{\NTT}{NTT Basic Research Laboratories, NTT Corporation, 3-1 Morinosato-Wakamiya, Atsugi, Kanagawa, 243-0198, Japan.}
\newcommand{\TQP}{NTT Theoretical Quantum Physics Center, NTT Corporation, 3-1 Morinosato-Wakamiya, Atsugi, Kanagawa 243-0198, Japan.}

\newcommand{\NII}{National Institute of Informatics, 2-1-2 Hitotsubashi, Chiyoda-ku, Tokyo 101-8430, Japan.}
\newcommand{\oxford}{Department of Materials, University of Oxford, OX1 3PH, United Kingdom}

\begin{document}
\title{
%Heisenberg limited quantum metrology under the effect of dephasing
%Heisenberg limited
Quantum metrology beyond the classical limit
under the effect of dephasing
}
 %%%%%%%%%%%%%%%%%%%%%%%%%%%%%%%%%%%%%%%%%%%%%%%%%%%%%%
 % \author{Munro-san?}                                %
 % \affiliation{                                      %
 % NTT Basic Research Laboratories, NTT Corporation,  %
 % Kanagawa, 243-0198, Japan                          %
 % }                                                  %
 %%%%%%%%%%%%%%%%%%%%%%%%%%%%%%%%%%%%%%%%%%%%%%%%%%%%%%

\author{Yuichiro Matsuzaki}
\email{matsuzaki.yuichiro@lab.ntt.co.jp} \affiliation{\NTT}\affiliation{\TQP}
%\author{Hiroshi Yamaguchi}		\affiliation{\NTT}
 \author{Simon Benjamin}       \affiliation{\oxford}
 \author{Shojun Nakayama} 	\affiliation{\NII}
 \author{Shiro Saito} 	\affiliation{\NTT}
 \author{William J. Munro}   \affiliation{\NTT}\affiliation{\TQP} \affiliation{\NII}
\begin{abstract}
 Quantum sensors \textcolor{black}{have} the potential to outperform their classical counterparts.
 %potentially outperform
 %be used to measure target fields with a
  %sensitivity beyond that of
 %existing
% any classical sensors.
For classical sensing, the uncertainty of the estimation of the target
 fields
 %decreases
 \textcolor{black}{scales inversely with the square root of the measurement time T.}
% $1/\sqrt{T}$ with  a
 %total measurement time.
 On the other hand, by using quantum \red{resources}, we can reduce
 \textcolor{black}{this} \textcolor{black}{scaling of the uncertainty with time}
 %uncertainty \textcolor{black}{in} time
 to \red{$1/T$}.
 %- the Heisenberg limit.
 However, as quantum states are susceptible to dephasing, it has not
 been clear whether we can achieve sensitivities
 \textcolor{black}{with a scaling of $1/T$
% beyond the classical shot-noise limit
 %the Heisenberg limit
 for a
 measurement time
 longer than the coherence time}. Here, we
 propose a scheme that estimates the amplitude of globally applied
 fields
 with the uncertainty of $1/T$
% with Heisenberg limited sensitivity
 %in principle
 for an arbitrary time
 scale under the effect of dephasing. We use one-way quantum
 computing
 based teleportation between qubits to prevent any increase in the correlation between
 the quantum state and its local environment
 from \textcolor{black}{building up}
 %increasing,
 and have shown that such a teleportation protocol can
 suppress the local dephasing while the information from the target
 fields keeps growing. Our method has the potential to realize a quantum sensor with a sensitivity far beyond that of any classical sensor.
\end{abstract}

\maketitle

\red{It is well known that two-level systems are attractive candidates}
with which to realize ultrasensitive sensors \red{as the frequency} of the qubit
can be shifted by coupling it to a target field. Such a frequency shift
induces a relative phase between the \red{qubits basis states which can
be simply measured in a Ramsey type experiment}. This method has been
used to measure magnetic fields, electric fields, and temperature.
%, and \red{even gravity waves}
\cite{budker2007optical,balasubramanian2008nanoscaleetal,maze2008nanoscaleetal,degen2016quantum}. With
the typical  \red{classical sensor} measurement devices (including
SQUID's \cite{simon1999local}, Hall sensors \cite{chang1992scanning},
and force sensors \cite{poggio2010force}),   the uncertainty \red{in}
the estimation of the target fields
\textcolor{black}{scales as}
%decreases only slowly as
\textcolor{black}{$1/\sqrt{T}$ with  a total measurement time $T$.} \red{This scaling is considered
classical
}
%known as the shot noise limit
\cite{huelga1997improvement}.
With a qubit-based sensor using a Ramsey type
measurement, \textcolor{black}{the readout signal is periodic against
the amplitude of the target fields.
So, unless the range of the target fields is known, the interaction time with the target fields should
be limited, which reduces the sensitivity. In this case, the
sensitivity decreases as $1/t\sqrt{N}$ by
performing $N$ repetitions with a short sensing time $t$. This
sensitivity can be rewritten as
$1/\sqrt{Tt}$ if fast qubit control is available.
%Under the assumption
%of negligible decoherence,
Although 
one could achieve the uncertainty with $1/T$
by setting $t=T$ with the knowledge of the target field range,
a dynamic range, which allows us to estimate the fields unambiguously,
becomes small due to the periodic structure of the readout signal.
% with the
% cost of a small dynamic range.
 %, which means 
%  unambiguous estimation of the fields can be realized only in a small
%  range of the magnetic fields.
 Fortunately, there is an ingenious way to improve the
 dynamic range by using a feedback control of the qubit
\cite{said2011nanoscale,higgins2007entanglement}.}
%the uncertainty
%could scale as
%  \textcolor{black}{decreases as $1/t\sqrt{N}$ by
% performing $N$ repetitions with a single sensing time $t$. This
% sensitivity can be rewritten as
% $1/\sqrt{Tt}$ if fast qubit control is available. Under the
% assumption of negligible decoherence,
% one can achieve the uncertainty with $1/T$ by setting $t=T$  with the
% cost of  a small dynamic range, which means 
%  unambiguous estimation of the fields can be realized only in a small
%  range. Fortunately, there is an ingenious way to improve the
%  dynamic range by using a feedback control of the qubit
% \cite{said2011nanoscale,higgins2007entanglement}.} 
% It is possible to achieve the uncertainty with $1/T$ by setting the
% sensing time $t$ as an order of $T$ if decoherence is negligible. Although a
% naive application of this strategy provides a small dynamic range due to
% a periodic modulation of the signal, there is an ingenious way to improve the
% dynamic range using the coherence properties of the qubit
%\cite{said2011nanoscale,higgins2007entanglement}.
% Quantum and especially qubit-based
% sensors can in principle decrease the uncertainty with time to
% $1/T$
%- the Heisenberg limit
%This
%can be achieved using the coherence properties of the qubit and
Actually, several experimental demonstrations
%using quantum feedback
have shown
%sub-shot noise
a sensitivity \textcolor{black}{that scales as $1/T$}
%measuring the amplitude of target fields
%with
\textcolor{black}{with the high-dynamic range}
%a high-dynamic range
\cite{higgins2007entanglement,waldherr2012high}.
However,
as quantum states are susceptible to decoherence, it has
\red{generally} been considered that such a scaling \textcolor{black}{$1/T$} can
only be realized if the measurement time $T$ is \red{much} shorter than
the coherence time
\cite{demkowicz2012elusive,said2011nanoscale}. \red{Recently several
approaches } have been proposed and demonstrated that use quantum error correction
\cite{gottesman2009introduction} and dynamical decoupling
\cite{viola1999dynamical,taylor2008high,de2011single} to circumvent this limitation. Using quantum error correction,
we can measure the amplitude of the target field
with an uncertainty scaling as $1/T$
%with the Heisenberg limited sensitivity
under the effect of specific decoherence such as bit
flip \red{errors}
\cite{dur2014improved,herrera2015quantum,arrad2014increasing,kessler2014quantum,cohen2016demonstration,unden2016quantum,aqec2017hl},
\red{while} dynamical decoupling makes it possible to estimate \red{the}
frequency of time-oscillating fields with sensitivity
beyond the
classical limit on time scale longer than the coherence time
\cite{asimon2017quantum,boss2017quantum}. However, there is currently no
%\red{currently}
known metrological scheme to achieve
\textcolor{black}{with an uncertainty of $1/T$}
%Heisenberg limited sensitivity
when measuring the amplitude of 
target fields \textcolor{black}{with dephasing}.
%under the effect of dephasing.

\red{In this letter,} we propose
%a scheme to realize
%a Heisenberg-limited sensing of
\textcolor{black}{a scheme for measuring}
\red{the} amplitude of target fields \textcolor{black}{with an uncertainty of $1/T$} under the effect of
dephasing. We will use a similar concept to the quantum Zeno effect
(QZE)
\cite{misra1977zeno,itano1990quantum,FacchiNakazatoPascazio01a}. For
shorter time scales than \red{the} correlation time of the environment
\red{$\tau_c$},  the interaction with the environment
induces a quadratic decay rate that is much slower than
\textcolor{black}{the typical} exponential
decay \cite{NakazatoNamikiPascazio01a}. \textcolor{black}{Frequent
measurements can be used} to reset the correlation with the environment
\textcolor{black}{and so keep this}
state in the initial quadratic decay region, which suppresses the
decoherence
\cite{misra1977zeno,itano1990quantum,FacchiNakazatoPascazio01a}. However,
if we naively apply the QZE to quantum metrology, the frequent
measurements freeze all the dynamics so that the quantum states cannot
acquire any information from the target fields. Instead, we use quantum
teleportation (QT)  \red{based on concepts taken from one-way quantum
computation}
\cite{Raussendorf:2001p368,barjaktarevic2005measurement,silva2007direct,olmschenk2009quantum,baur2012benchmarking}
to reset the correlation between the system and the environment
\cite{averin2016suppression}.
If we \red{transfer} the quantum states to a new site, we can prevent
any increase in
the correlation between the system and environment
in the previous site, and the quantum state are
\red{then only} affected by a slow quadratic decay due to
the local environment in the new site. This noise suppression
\textcolor{black}{with a qubit motion using a concept drawn from QT}
has been
proposed and demonstrated by using superconducting qubits
\cite{averin2016suppression}.
The crucial idea in this paper is to use this one-qubit
teleportation-based noise suppression for quantum metrology.
Interestingly, although the QT protocol eliminates the deterioration
effect caused by the dephasing from the local environment,
we can accumulate the phase information from the global target fields
during this protocol.
We have shown that, as long as nearly perfect QT is available, we can
achieve \textcolor{black}{a sensor with the uncertainty scaling $1/T$}
%the Heisenberg limit sensing
%under the effect of local
\textcolor{black}{with}
dephasing.
Moreover, we have found that, even when the QT is moderately noisy, the
sensitivity of our protocol is superior to that of the standard Ramsey measurement.

\textcolor{black}{{\it{Noise and its suppression.--}}}
Our system and the environment \textcolor{black}{in this situation} can be described by a Hamiltonian of the
form $H=H_{\rm{S}}+H_{\rm{I}}+H_{\rm{E}}$
\cite{hornberger2009introduction}
%  \begin{eqnarray}
%   H_{\rm{S}}&=&\sum_{j=1}^{L}\frac{\omega }{2}\sigma ^{(j)}_z\otimes \openone^{(j)}_{\rm{E}}, \;\;\;\;
%   H_{\rm{I}}=\sum_{j=1}^{L}\lambda \sigma ^{(j)}_z\otimes B_j,
%    \nonumber \\
% &\;&\;\;\;\;\;\;\;\;\;\;\;\;
%   H_{\rm{E}}=\sum_{j=1}^{L}\openone^{(j)}_{\rm{S}} \otimes C_j,
%  \end{eqnarray}
 where
 $H_{\rm{S}}=
 %\hbar
 \sum_{j=1}^{L}\frac{\omega }{2}\sigma ^{(j)}_z\otimes \openone^{(j)}_{\rm{E}}$ ($H_{\rm{E}}=\sum_{j=1}^{L}\openone^{(j)}_{\rm{S}} \otimes C_j$)
 denotes \red{the} system
(environmental) Hamiltonian \red{while}
$H_{\rm{I}}=
%\hbar
\sum_{j=1}^{L}\lambda \sigma ^{(j)}_z\otimes B_j$ denotes the
interaction between the system and
the environment.
%We set $\hbar=1$.
 \red{Here $\sigma ^{(j)}_z$ is the usual Pauli Z operator
of the $j$-th qubit with frequency $\omega $, while} $B_j$ and $C_j$
denote the environmental operator at \red{that} $j$-th
 site. $\hat{\openone}^{(j)}_{\rm{S}}$ ($\hat{\openone}^{(j)}_{\rm{E}}$)
 denotes an identity operator for the system
 (environment). Furthermore, we set $\hbar=1$. In an interaction picture, we have
 $ H_{\rm{I}}(t)=
 %hbar
 \lambda \sum_{j=1}^{L}\sigma ^{(j)}_z\otimes
 \tilde{B}_j(t)$ where
 $\tilde{B}_j(t)=e^{iH_{\rm{E}}t      }B_je^{-iH_{\rm{E}}t }$.
 The \textcolor{black}{separable} initial state is given as $\rho (0)=  \bigotimes _{j=1}^L(\rho
 ^{(j)}_{\rm{S}}(0) \otimes \rho^{(j)}_{\rm{E}})$.
 where we have assumed $\rho^{(j)}_{\rm{E}}$ is in thermal
 equilibrium ($[\rho^{(j)}_{\rm{E}},H_{\rm{E}}]=0$) and
 \red{our noise  is} non-biased (${\rm{Tr}}[\rho^{(j)}_{\rm{E}}B_j] =0$) for all $j$. If the initial
 state is separable,
 we consider
 the first site by tracing out the others. %Solving
 Solving Schrodinger's equation gives
% let us obtain
 %for a time $\tau $,
% we obtain
 \textcolor{black}{
\begin{eqnarray}
 \rho ^{(1)}_{\rm{I}}(\tau )\simeq \rho ^{(1)}(0)-i\lambda
  \int_{0}^{\tau }dt' [\sigma ^{(1)}_z\otimes
  \tilde{B}_1(t'),\rho ^{(1)}(0)]\nonumber \\
 -\lambda ^2\int_{0}^{\tau }\int_{0}^{t'}dt'dt'' [\sigma ^{(1)}_z\otimes
  \tilde{B}_1(t'),[\sigma ^{(1)}_z\otimes
  \tilde{B}_1(t''),\rho ^{(1)}(0)]]\nonumber
\end{eqnarray}}using a second order perturbation expansion in $\lambda $
%\textcolor{black}{where $\tau $ denotes a time}
\cite{hornberger2009introduction}. Tracing out the environment, we have
\red{
\begin{eqnarray}
\rho ^{(1)}_{\rm{S}}(\tau )&\simeq& \rho ^{(1)}_{\rm{S}}(0) \nonumber \\
&& -\lambda ^2\int_{0}^{\tau }\int_{0}^{t'}dt'dt''  C^{(1)}_{t'-t''} \left[\hat{\sigma }^{(1)}_z,\left[\hat{\sigma }^{(1)}_z,\rho ^{(1)}_{\rm{S}}(0)\right]\right]
  \nonumber
\end{eqnarray}}
where we define \red{the} correlation function of the
environment as
\textcolor{black}{
$C^{(1)}_{t'-t''}\equiv\frac{1}{2}
{\rm{Tr}}[(\tilde{B}_1(t')\tilde{B}_1(t'')+\tilde{B}_1(t'')\tilde{B}_1(t'))\rho^{(1)}_{\rm{E}}]$.}
If
we are interested in a time scale
%\textcolor{black}{the relevant time scale is}
much shorter than the
%\textcolor{black}{environmental correlation time}.
correlation time of the environment,
we can approximate the correlation function as $C^{(1)}_{t'-t''}\simeq
C^{(1)}_{0}$.
\textcolor{black}{For most solid state systems, this is readily satisfied as the environment correlation time is much longer than the coherence time of the qubit}
% For most of the solid state systems,
% the correlation time
% is much longer than the
% coherence time of the qubit
\cite{de2010universal,YoshiharaHarrabiNiskanenNakamura01a,KakuyanagiMenoSaitoNakanoSembaTakayanagiDeppeShnirman01a,kondo2016using},
and so this condition is readily satisfied for many systems.
\textcolor{black}{In such a case}
$\rho ^{(1)}_{\rm{S}}(\tau )\simeq (1-\epsilon _\tau )U_{1,\tau
}\rho^{(1)}_{\rm{S}}(0)U_{1,\tau }^{\dagger }+\epsilon _\tau \hat{\sigma
}^{(1)}_z  U_{1,\tau }\rho ^{(1)}_{\rm{S}}(0) U_{1,\tau}^{\dagger
}\sigma ^{(1)}_z $
\textcolor{black}{with $U_{j,\tau}=e^{-{i\omega \tau}\hat{\sigma
}^{(j)}_z/2}$ specifying the unitary operator for a site $j$ and
$\epsilon _{\tau }=\lambda ^2 C_0 \tau ^2$ denoting the error rate for $\lambda ^2 C_0 \tau^2 \ll 1$.}
% where \textcolor{black}{$\epsilon _{\tau }=\lambda ^2 C_0 \tau ^2$
% denotes an error rate for $\lambda ^2 C_0 (\tau)^2 \ll 1$} and
% $U_{j,\tau}=e^{-{i\omega \tau}\hat{\sigma }^{(j)}_z/2}$
% denotes a unitary operator at a site $j$.
 Since the error rate has a quadratic form
in time $t$,
the decoherence effect is \red{negligible for short time scales $t\ll
{1}/{\lambda \sqrt{C_0}}$}.
%which 
This has been discussed in the field of the QZE
\cite{misra1977zeno,itano1990quantum,FacchiNakazatoPascazio01a}.
On the other hand, if we consider longer time scales of  $t>
{1}/{\lambda \sqrt{C_0}}$ with the same environment,
error accumulation will destroy the quantum coherence of the qubit.

Let us now describe the noise suppression technique using QT. It
begins
 %but 
 \textcolor{black}{with}
free evolution of the qubit for a time $\tau=t/n$
(where $t$ is the total time and $n$ is the number of times QT is \textcolor{black}{to be
performed}). After this, QT \textcolor{black}{transports} $\rho ^{(1)}_{\rm{S}}$ to 
site $2$. The quantum state starts interacting with a new
local environment described by a density matrix $\rho
^{(2)}_{\rm{E}}$. The error rate will be suppressed due to the
quadratic decay \cite{averin2016suppression}. \red{Performing QT $n$
times (each time to a fresh qubit) yields}
$ \rho ^{(n)}_{\rm{S}}(t)\simeq (1-\frac{\lambda ^2 C_0
t^2}{n})U_{n,t}\rho_{\rm{S}}(0)U^{\dagger}_{n,t}
+\frac{\lambda ^2 C_0 t^2}{n}\hat{\sigma }_z U^{\dagger}_{n,t}
\rho_{\rm{S}}(0) U^{\dagger}_{n,t}\sigma _z $
 at  site $n$. \red{For a large $n$ this approaches the} pure state
 $\rho _{\rm{S}}(t) \simeq U_{n,t} \rho_{\rm{S}}(0) U^{\dagger}_{n,t}$,
 and so our approach can suppress dephasing.

 \textcolor{black}{
 {\it{Definition of parameters.--}}
Here, we discuss the key parameters that we will use in our scheme. 
%  We define $T$, $L$, $t$, $M$, and $n$ as the total sensing time, the
% total number of probe qubits,
% the interaction time with the field, the size of the entangled state, and the number of the QTs during the
% interaction time, respectively.
\textcolor{black}{We define $L$ and $M$ as the total number of the probe
 qubits and the size of the entangled state. In our scheme, there are three time scales, $\tau \ll  t \ll  T$. The interaction time
 between teleportations is denoted $\tau $. This interaction  is
 repeated $n$ times between the state preparation and measurement, giving a
 total time denoted $t=n\tau $. The whole procedure including
 preparation and measurement is repeated $N$ times, giving a total
 interaction time $T$.
}
% \textcolor{black}{ We define $L$, $M$, and $n$ as  the
% total number of probe qubits and the size of the entangled state, and the number of the QTs between the state preparation and readout. We define $T$ as a total sensing
% time that includes the repetitions to decrease the uncertainty of the
% estimation, while we define $t$ as an interaction time between the probe qubits
% and target field in a single measurement cycle. These lead us to define
% $\tau=t/n$ as an evolution time of the probe qubit between the QTs.}
As regards the dephasing model, although our general approach described above uses a  perturbative
analysis typically valid for a short time scale,
we need to examine the dynamics of our system for arbitrary
time scales.  So we will consider a more specific noise model that is given
by
$\hat{\mathcal{E}}_j(\rho ^{(j)}_{\rm{S}}(\tau ))= \frac{1+e^{-\gamma ^2 \tau^2}}{2}U_{j,\tau
}\rho^{(j)}_{\rm{S}}(0)U_{j,\tau}^{\dagger}
+\frac{1-e^{-\gamma ^2 \tau^2}}{2} \hat{\sigma }^{(j)}_z
U_{j,\tau}\rho^{(j)}_{\rm{S}}(0)U_{j,\tau}^{\dagger} \sigma ^{(j)}_z $
at the site $j$
during the evolution for a time $\tau=t/n$ (with $\gamma$ representing the
dephasing rate). This model is consistent with the general
%short time scale noise model
results
described above when we choose \textcolor{black}{ $\gamma
=\sqrt{2\lambda ^2 C_0}$ (See the supplementary materials)}. Typical dephasing models
\cite{PSE,de2010universal,KakuyanagiMenoSaitoNakanoSembaTakayanagiDeppeShnirman01a,YoshiharaHarrabiNiskanenNakamura01a}
show this behavior if the correlation time of the environment is much
longer than the dephasing time.
\textcolor{black}{If we consider a state 
$\rho $ composed of $M$ qubits, the noise channel during the time
evolution of $\tau $ is described as
$\hat{\mathcal{E}}_1\hat{\mathcal{E}}_2\cdots \hat{\mathcal{E}}_M(\rho
)$. }  
%During the interaction with the target fields,
\textcolor{black}{We
%\red{could}
consider GHZ states
$|\psi ^{{\rm{(GHZ)}}}\rangle =\frac{1}{\sqrt{2}}\left[\bigotimes
_{j=1}^{M}|0\rangle_j+ \bigotimes _{j=1}^{M}|1\rangle
_j\right]$ as a metrological resource \cite{huelga1997improvement,shaji2007qubit,bohmann2015entanglement}.
%with a size of $M$ qubits.
For a given $L$ qubits, we create GHZ states with a size of $M$ qubits, and the
number of the GHZ states is ${L}/{M}$.
% In our previous ideal QT analysis,
% the uncertainty of the estimation monotonically decreases as the number
% of the QT increases.
% However in realistic situations, errors caused during the QT
% operations will limit our achievable sensitivity and
% there will be an optimal number of QT's that we can perform.
 %Moreover, we will use QT in our sensing scheme, and consider an
 %imperfect QT as follows.
% If a perfect QT is available, the uncertainty of the estimation would monotonically decrease as the number
% of the QT increases.
 In realistic situations, there will be errors caused during the QT
operations,
%could exist
%will not be negligible
%will limit our achievable sensitivity
%and there will be an optimal number of QT's that we can perform,
and so we consider an
 imperfect QT.
 If we teleport a state
 $\rho _1$ from $j=1,2,\cdots ,M$ sites to $j'=1+M,2+M,\cdots ,2M$
 sites, we obtain a state of
 $\rho '_{2}=(1-p )^M\rho _{2}+(1-(1-p )^M)   \rho
 ^{(\rm{error})}_{2}$ where $p$ is the error rate on a single
 qubit, $\rho _{2}$ is the ideal state
 (that we could obtain by a perfect QT), and $\rho ^{(\rm{error})}_{2}
 =\frac{1}{2}(\bigotimes_{j=1+M}^{2M}|0\rangle _j \langle 0| +\bigotimes
 _{j=1+M}^{2M}|1\rangle_j \langle 1|)$ is the dephased state.
 %a decohered state.
%  Here, we assume that any error on a single qubit
% makes the probe state into $\rho ^{(\rm{error})}_2$.
}
}

% \begin{figure*}[t!] 
% \begin{center}
% \includegraphics[width=0.9\linewidth]{table.eps} 
% \caption{
% \textcolor{black}{Performance of our teleportation based scheme
%  with $L$ qubits for a given time $T$. The uncertainty of the standard Ramsey scheme is given as 
%  $\delta
%  \omega_{{\rm{R}}}=e^{{1}/{4}}\sqrt{\gamma}/\sqrt{TL}$. With imperfect
%  quantum teleportation that has an error rate of $p$, we can
%  \textcolor{black}{achieve a sensitivity}
%  %beat the classical limit
%  scaled as $1/T$ using separable states (entangled states with a size of
%  $M$) for a
%  short time scale of $T\ll 1/\sqrt{p}\gamma $ ($T\leq 1/4\sqrt{p}\gamma M$).
%  % Under the realistic assumption that we use a high fidelity quantum
% %  teleportation
%  For a longer time scale, if accurate quantum teleportation is available
%  ($p\ll 1$), the sensitivity of our scheme can be
%  still better than that of the standard Ramsey scheme.
%  }}
%  \label{table}
% \end{center}
%   \end{figure*}

  \begin{table*}
\begin{center}\textcolor{black}{
    \begin{tabular}{| l | l | l | l | p{4.8cm} |}
    \hline
     &General form of sensitivity &\ \ \ \ \  Perfect QT & \ \ \ \
     %Imperfect QT for a short T
     Short T imperfect QT
     & \ \ \ \
     %Imperfect QT for a long T
     Long T imperfect QT
     \\ \hline
      Entangled sensor  & \scalebox{1.1}{$ \delta \omega 
  \simeq \frac{{\rm{Exp}}[M\gamma^2t^2/n]}{(1-p )^{M(n-1)}\sqrt{MLTt}}$}
         &\scalebox{1.1}{$\delta \omega \simeq
             \frac{{\rm{Exp}[1/4]}}{T\sqrt{ML}}$} &  \scalebox{1.1}{$\delta \omega \simeq
             \frac{{\rm{Exp}[1/4]}}{T\sqrt{ML}}$} for $T\ll  \frac{1}{\sqrt{p}\gamma M}$ &
                      \scalebox{1.1}{$\delta
 \omega \simeq 2^{\frac{3}{4}}\sqrt{\frac{e\sqrt{p
 }\gamma}{TL}}$} for $T\gg   \frac{1}{\sqrt{p}\gamma M}$
                     \\ \hline
     Separable  sensor&
         \scalebox{1.1}{$ \delta \omega 
  \simeq \frac{{\rm{Exp}}[\gamma^2t^2/n]}{(1-p )^{(n-1)}\sqrt{LTt}}$} &
             \scalebox{1.1}{$\delta \omega \simeq
             \frac{{\rm{Exp}[1/4]}}{T\sqrt{L}}$} &
                  \scalebox{1.1}{$\delta \omega \simeq
             \frac{{\rm{Exp}[1/4]}}{T\sqrt{L}}$} for $T\ll \frac{1}{\sqrt{p}\gamma} $
                 & \scalebox{1.1}{$\delta
 \omega\simeq 2\sqrt{\frac{e\sqrt{p
 }\gamma}{TL}}$} for $T\gg \frac{1}{\sqrt{p}\gamma} $ \\ 
    \hline
    \end{tabular}}
 \caption{
\textcolor{black}{Performance of our teleportation based scheme
 with $L$ qubits for a given time $T$. \textcolor{black}{Except with the
 general form, we show
 optimized sensitivity by choosing a suitable interaction time ($t$)  and
 the QT number ($n$).}
 The uncertainty of the standard Ramsey scheme is given as 
 $\delta
 \omega_{{\rm{R}}}=e^{{1}/{4}}\sqrt{\gamma}/\sqrt{TL}$. With imperfect
 quantum teleportation that has an error rate of $p$, we can
 \textcolor{black}{achieve a sensitivity}
 %beat the classical limit
 \textcolor{black}{scaling} as $1/T$ using separable states (entangled states with a size
 $M$) for a
 short time scale of $T\ll 1/\sqrt{p}\gamma $ ($T\ll 1/\sqrt{p}\gamma M$).
 % Under the realistic assumption that we use a high fidelity quantum
%  teleportation
 For a longer time scale, if accurate quantum teleportation is available
 ($p\ll 1$), the sensitivity of our scheme can be
 still better than the sensitivity of the standard Ramsey scheme.
  \textcolor{black}{
 It is worth mentioning that, for the general form, perfect QT, and
 short T imperfect QT, the sensitivity of the separable sensor can be simply
 obtained by setting $M=1$ in that of the entangled sensor.}
 }}
 \label{table}
\end{center}
   \end{table*}

{\it{Quantum metrology with QT.--}}
% A natural question we have not addressed so far is whether
% entanglement improves our 
% sensing with QT.
%Let us now
Here, we focus
%our attention
on
using the QT scheme to enable quantum
metrology with an uncertainty scaling as $1/T$.
Consider the situation in which the qubit frequency $\omega$  is
shifted depending on the amplitude of the target fields,
and so measurement of the qubit's frequency shift allows us to
infer the amplitude of the target field.
Such a qubit frequency shift is estimated from the relative
phase between quantum states.
The key idea
is to use the
QT in  a ring arrangement with $2L$ qubits where each qubit has a tunable interaction with
another qubit.
%with it's nearest neighbor qubits
%\textcolor{black}{We illustrate our scheme with separable states in
%Fig. \ref{schematic}.}
Half of the qubits are used to probe the target fields while
the \textcolor{black}{remaining} qubits are used as an ancilla for QT.
%(probe qubits are located between two ancillary qubits).
The QT is
accomplished by implementing a control-phase gate between a probe qubit  and
an ancilla qubit,
followed by a $\hat{\sigma }_x$ measurement  on the probe qubit
(and single qubit corrections depending on the measurement result).
This QT approach has been widely used in one-way quantum computation \cite{Raussendorf:2001p368,browne-2006-}.

{\it{Scheme with entanglement.--}}
Our scheme for measuring the amplitude of the target fields is as follows:
\textcolor{black}{First, we prepare GHZ states of $\bigotimes _{k=0}^{\frac{L}{M}-1}|\psi
^{{\rm{(GHZ)}}}_k\rangle$ between the probe qubits where $|\psi ^{{\rm{(GHZ)}}}_k\rangle =\frac{1}{\sqrt{2}}\left[\bigotimes
_{j=1+2kM}^{M+2kM}|0\rangle_j+ \bigotimes _{j=1+2kM}^{M+2kM}|1\rangle
_j\right]$ for $k=0,1,\cdots ,(\frac{L}{M}-1)$, while the other qubits
(which we call ancillary qubits) are prepared in $|0\rangle $. 
%located at the site $2j-1$ ($j=1,2,\cdots ,L$).
Second we
let the  state evolve for time $\tau ={t}/{n}$ and then teleport
 the state of the probe qubit at the site $j$
%$j=1+2kM, 2+2kM, \cdots , M+2kM$ (for $k=0,1,\cdot ,(\frac{L}{M}-1)$)
to another site $j+M$}.
%$j=1+(2k+1)M, 2+(2k+1)M, \cdots , M+(2k+1)M$ (for $k=0,1,\cdot ,(\frac{L}{M}-1)$) 
We assume that our gate operations are much faster than
$\tau$. 
Third,
%we then in step 3
we repeat the second step $(n-1)$
times, \red{while in the fourth step}
we let this state evolve for time $\tau =\frac{t}{n}$, and readout
the states.
Finally, we repeat these steps $N$ times during time $T$
where $N\simeq {T}/{t}$ is the repetition number.

We derive the sensitivity using imperfect QT
and entanglement with general conditions, and subsequently discuss special cases.
% We
% create a GHZ state
% composed of $M$ qubits,
%For a given $L$ qubits, we create GHZ states with this size,
%and the number of the GHZ states is ${L}/{M}$.
% We can use these GHZ states to detect the fields with
% QT between the GHZ states.
By letting the GHZ states $|\psi^{{\rm{(GHZ)}}}_k\rangle$
evolve with low-frequency dephasing for time $\tau$, we have
\textcolor{black}{
\begin{eqnarray}
 \rho _{k}(\tau )=\frac{1}{2}(\bigotimes _{j=1+2kM}^{M+2kM}|0\rangle _j \langle  0|+\bigotimes _{j=1+2kM}^{M+2kM}|1\rangle _j \langle 0|e^{-iM\omega
 \tau-M\gamma ^2t^2}\nonumber \\
 +\bigotimes _{j=1+2kM}^{M+2kM}|0\rangle _j \langle 1|e^{iM\omega
  \tau-M\gamma ^2\tau^2}+\bigotimes _{j=1+2kM}^{M+2kM}|1\rangle _{j}
  \langle  1|)\ \ \ \ \ 
  \nonumber 
\end{eqnarray}}
for $k=0,1,\cdots ,(\frac{L}{M}-1)$ where $\gamma$ denotes the dephasing
rate for a single qubit.
%It is worth mentioning that,
\textcolor{black}{If
we use the QT many times, we can suppress the low-frequency dephasing
by employing the mechanism that we described before.}
To readout the GHZ states, we measure a projection operator defined by
$\hat{\mathcal{P}}^{(k)}_{\pm } =|\psi ^{(\pm )}_{\perp}\rangle _k
\langle \psi ^{(\pm )}_{\perp}|$
 where \textcolor{black}{$|\psi ^{(\pm )}_{\perp}\rangle _k=\frac{1}{\sqrt{2}}\bigotimes
 _{j=1+2kM}^{M+2kM}|0\rangle _j\pm  i\frac{1}{\sqrt{2}}\bigotimes
 _{j=1+2kM}^{M+2kM}|1\rangle _j$.}
%  We consider an imperfect QT as follows. If we teleport a state of
%  $\rho _1$ from $j=1,2,\cdots ,M$ sites to $j'=1+M,2+M,\cdots ,2M$
%  sites, we obtain a state of
%  $\rho '_{2}=(1-p )^M\rho _{2}+(1-(1-p )^M)   \rho
%  ^{(\rm{error})}_{2}$ where $p$ denotes \red{the} error rate on a single
%  qubit, $\rho _{2}$ denotes the ideal state
%  (that we could obtain by a perfect QT), and $\rho ^{(\rm{error})}_{2}
%  =\frac{1}{2}(\bigotimes_{j=1+M}^{2M}|0\rangle _j \langle 0| +\bigotimes
%  _{j=1+M}^{2M}|1\rangle_j \langle 1|)$ denotes a decohered state.
%  Here, we assume that any error on a single qubit
% makes the probe state into $\rho ^{(\rm{error})}_2$
% (\red{The $M=1$ corresponds to our previous case where we used separable
% states}).
 We can then estimate the sensitivity in this situation as
\begin{eqnarray}
 \delta \omega ^{({\rm{GHZ}})}_{n,t,M}=\frac{\sqrt{\langle \delta
  \hat{\mathcal{P}}_{\pm } \delta \hat{\mathcal{P}}_{\pm }\rangle }}{|{d
  \langle \hat{\mathcal{P}}_{\pm } \rangle }/{d\omega
  }|\sqrt{N}}
  =\frac{e^{{M\gamma^2t^2}/{n}}}{(1-p )^{M(n-1)}\sqrt{MLTt}}\ \label{generalsensitivity}
\end{eqnarray}
where  $\delta \hat{\mathcal{P}}= \hat{\mathcal{P}} - \langle
 \hat{\mathcal{P}}\rangle$ and \red{$N\simeq {TL}/{tM}$}.
  \textcolor{black}{From this general formula, we can derive many special
 cases by substituting parameters, which we will describe below. Also,
 a summary of these
 results is shown in Table \ref{table}.}
 %It is
 %worth mentioning that,
 By setting
 $n=1$, we can reproduce the results discussed in
 \cite{schaffry2010quantum,jones2009magnetic,matsuzaki2011magnetic,chin2012quantum}
 for an entanglement based sensor with low-frequency dephasing.
% Our general approach described above has used a  perturbative
% analysis typically valid on a short time scale.
% However we need to examine the dynamics of our system for arbitrary
% time scales, and

For the perfect QT  ($p=0$), we
 achieve
  the Heisenberg limit
  $\delta \omega ^{\rm{(GHZ)}}_{n,t_{\rm{opt}},M}=e^{1/4}/\sqrt{c}LT$
  when we set $t=T$, $M=cL$, and $n=4M\gamma^2T^2$ where $c$ denotes a
  constant number.
 \textcolor{black}{
 However, since the entangled state can be teleported to the
  original site where the entangled state previously interacted
with the environment, correlated error may be induced
due to the environmental memory effect.
This could happen for $n\geq \tilde{n}_{\rm{en}} $ where
 $\tilde{n}_{\rm{en}}$ denotes the maximum teleportation number of
the teleportation without the entangled state being teleported back
to the original site.   In this case, we have $\tilde{n}_{\rm{en}}=\frac{2L}{M}-1$.}
%   as the maximum number of the teleportation without the entangled state
% being teleported back to the original site.
%Since
The typical environment has a finite
 correlation time $\tau _c$.
 %such a correlated effect becomes negligible if
%To consider the error as independent,
%we need
%a condition of
Unless the condition
  $\tilde{n}_{\rm{en}}\tau \gg \tau _c \Leftrightarrow  c^2L\gamma ^2
  T\tau _c\ll 1$ is satisfied, the error could be correlated (See the supplementary
 materials). \textcolor{black}{Also,  to
  observe the quadratic decay, we need a condition of $\tau _c\gg
  t/n$. This means that the correlation time should satisfy these two
  conflicting conditions.}
  %to make the error independent
  %is satisfied
  %is necessary
  %This means that,
  %This means that,
  So, although we
  observe the Heisenberg limit scaling for a small
  $L$, the correlated error would begin to hinder the Heisenberg limit as
  we increase the size of the entangled state.
 
\textcolor{black}{A natural question is what happens if our QT is imperfect and so we consider that here. }
%  We consider the performance of the entanglement scheme with
 %imperfect QT.
 \textcolor{black}{For short times $T\ll  1/\sqrt{p}\gamma M$,
 the error due to the QT is negligible, and so we obtain the same
 results as in the perfect QT case by setting $t=T$ and $n=4M\gamma^2T^2$.}
%  we obtain $ \delta
%  \omega ^{({\rm{GHZ}})}_{n,t,M} \simeq e^{1/4+M\gamma
%  ^2T^2/n}/T\sqrt{ML}\leq e^{1/2}/T\sqrt{ML}$ when $t=T$ and $n=1/4Mp$. The condition for the error to be independent ($\tilde{n}_{\rm{en}}\tau
%  \geq \tau _c$) is written as $LTp\gg \tau _c$, which can be satisfied
%  for a large $L$.
 %\textcolor{blue}{We consider a scaling of the uncertainty for a longer
 %$T$.}
 \textcolor{black}{In quantum metrology, another interesting regime that is quite often
considered is the scaling law in the limit of long T (much greater than
the coherence time of the system). We consider this here.}
 %\textcolor{black}{For such a limit of long $T$,}
    We can minimize the uncertainty with $t^{(\rm{en})}_{\rm{opt}}=\frac{\sqrt{n/M}}{2\gamma }$ \red{to} obtain
$\delta
 \omega^{({\rm{GHZ}})}_{n,t_{\rm{opt}}}=\frac{\sqrt{2}e^{{1}/{4}}\sqrt{{\gamma}/{\sqrt{Mn}}}}{(1-p
 )^{M(n-1)}\sqrt{LT}}$ for $p>0$
 and $n> 1$. Furthermore, with $M_{\rm{opt}}=-{1}/{4\log (1-p)}\simeq {1}/{4p}$
 and $n^{(\rm{en})}_{\rm{opt}}=2$,
 %we minimize
 the uncertainty can be minimized as $\delta
 \omega^{(\rm{GHZ})}_{\rm{opt}}=2^{\frac{3}{4}}\sqrt{\frac{e\sqrt{p
 }\gamma}{LT}}$.
%  \textcolor{blue}{The condition for the error to be independent ($\tilde{n}_{\rm{en}}\tau
%  \geq \tau _c$) is satisfied just by increasing the number of the
%  qubits for $\tilde{n}_{\rm{en}}=L/M_{\rm{opt}}$ and $\tau
%     =t^{(\rm{en})}_{\rm{opt}}/n^{(\rm{en})}_{\rm{opt}}$.}
    In this case, the
 condition for the independent error
 ($\tilde{n}_{\rm{en}}\tau
 \geq \tau _c$) is written as $Lp^{3/2}/\gamma \gg \tau _c$, and is
 satisfied for a large $L$.
% The sensitivity has a constant factor improvement over our scheme
% with separable states.
 %,  \textcolor{black}{which is consistent with the fact that shows a fragility of an entanglement sensor}
 %\cite{huelga1997improvement,shaji2007qubit,bohmann2015entanglement},
%. This is consistent with
%\textcolor{black}{which is }
%the fact that small noise sometimes
% makes entanglement sensors almost equivalent with separable sensors

 \begin{figure}[h!] 
\begin{center}
\includegraphics[width=0.85\linewidth]{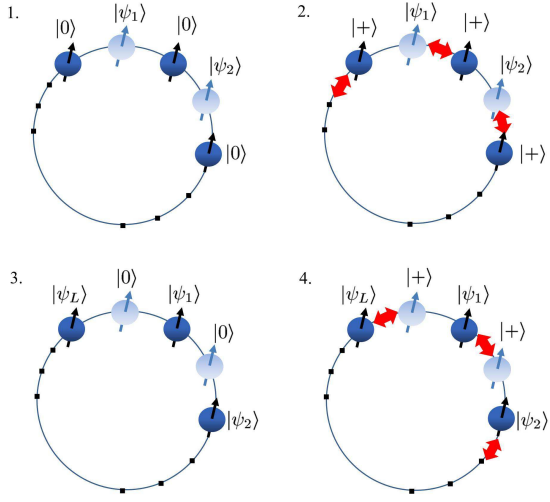} 
\caption{Schematic illustration of $2L$ qubits in a ring structure
 to measure globally applied fields \textcolor{black}{with an uncertainty
 scaling as $1/T$ when we use separable states.}
 %at the Heisenberg limit.
 Half of the qubits contain information about the target fields as a
 probe \red{while the remaining half  are used as ancillary qubits
 for the qubit teleportation}.
 % By performing a controlled phase gate
%  between the probe qubit and ancillary qubit, single qubit
%  measurements on the probe qubit with feed-forward operations
%  depending on the measurement results,
 \textcolor{black}{With a controlled phase gate and measurement feedforward operations,}
 we can teleport
 a quantum state $|\psi \rangle $ from the original site to the right
 neighboring site  \cite{browne-2006-,Raussendorf:2001p368}.
 After the teleportation, \red{the measured qubit becomes the new
 ancilla which we} initialize into $|0\rangle$.
 %By implementing the teleportation frequently, dephasing from the local environment can be suppressed while quantum states keep accumulating a phase information from the target
 %global fields.
 }
 \label{schematic}
\end{center}
\end{figure}

{\it{Scheme with separable states.--}}
\textcolor{black}{
%We will explain
Now we explore a possibly more practical scheme with
separable states, as shown in Fig. \ref{schematic}.
%First, we prepare a probe state of
We begin by preparing a probe state of
$\bigotimes _{j=1}^L|+\rangle
_{2j-1}$  located at the site $2j-1$ ($j=1,2,\cdots ,L$). Then we
let the  state evolve for a time $\tau ={t}/{n}$ and  teleport
 the state of the probe qubit to the next site using the
ancillary qubit.
%\red{Third we then in step 3 repeat the} second
We repeat this
step $(n-1)$
times
before we finish by allowing our
%\red{while in the fourth step}
%we let this
state to evolve for time $\tau =\frac{t}{n}$, and reading out
the state by measuring $\hat{M}_y=\sum_{j=1}^{L}\hat{\sigma }_y^{(j)}$.
%Finally,
We repeat these steps $N$ times during the measurement time $T$
where $N\simeq {T}/{t}$ is the repetition number.}
\textcolor{black}{We can calculate the sensitivity for this scheme by substituting $M=1$ in Eq. \ref{generalsensitivity}.}
% When $n$ is even, the density matrix before the readout is described as
% $\rho _{{\rm S}}(t)\simeq \bigotimes _{j=1}^L \rho
% ^{(2j-1)}_{\rm{S}}(t)$
% where $\rho ^{(2j-1)}_{\rm{S}}(t) =\frac{1}{2}(\openone
% _{2j-1}+|1\rangle _{2j-1}\langle 0|e^{-i\omega t -{\gamma
% ^2t^2}/{n}}+|0\rangle _{2j-1}\langle 1|e^{i\omega t -{\gamma ^2t^2}/{n}}
% )$. Here, \red{${\gamma }/{\sqrt{n}}$} can be interpreted as an \red{QT}
% improved coherence time.
% If $n$ is odd, we obtain the same density matrix for the probe qubits at
% a site $2j$ $(j=1,2,\cdots ,L)$.
% \red{In the situation that  the target field to be sensed is weak ($L\omega T\ll 1$) we can estimate} the uncertainty in our \red{estimator as}
%  \begin{eqnarray}
%   \delta \omega _{n,t}= \frac{\sqrt{\langle \delta \hat{M}_y \delta \hat{M}_y\rangle }}{|\frac{d \langle \hat{M}_y\rangle }{d\omega}|} \frac{1}{\sqrt{N}} \simeq \frac{e^{{\gamma ^2t^2}/{n}}}{\sqrt{LTt}}
% \end{eqnarray}
%  where $\delta \hat{M}_y = \hat{M}_y -\langle \hat{M}_y \rangle
%  $.
\textcolor{black}{For an ideal QT, by setting $t=T$ and $p=0$}, we obtain
 \textcolor{black}{$\delta \omega _{n,T}\simeq
 {e^{1/4}}/{T\sqrt{L}}$ for $n=4\gamma ^2T^2$},
 \red{and so we can achieve $1/T$ scaling.}
 %we can achieve the Heisenberg limit.
% \textcolor{black}{we can beat the classical shot-noise limit.}
% If the qubit state is
% teleported back to the original site where the qubit interacted with the
% environment before, correlated
% error may be induced due to the
% environmental memory effect. This could happen
For $n\geq \tilde{n}$,  a correlated
 error may be induced due to the
 memory effect where $\tilde{n}=2L-1$
denotes the maximum number of teleportations without the qubit state
being teleported back to the original site.
 Fortunately, since the typical environment has a finite
 correlation time $\tau _c$, such a correlation effect becomes negligible for a large number of qubits to satisfy $\tilde {n}\tau \gg 
 \tau _c\Leftrightarrow L\gg
\gamma ^2 T\tau _c $ (See the supplementary materials).
%We discuss these in
%(See the supplementary materials).

We now analyze how imperfect QT affects the performance of our
sensing scheme.
\textcolor{black}{We can calculate the sensitivity by substituting
$M=1$ and $p>0 $ with Eq. \ref{generalsensitivity}.}
% \red{We will consider a simple error model, a depolarization channel}
% that makes the state completely mixed with a probability
% $p$ after the QT operation.
% In such a case, after the \red{ imperfect QT
% is performed on say the $j$ th site,   the state $\rho _j$ evolves}
% to $(1-p)\rho _{j+1} +p {\openone_{j+1}}/{2}$. The uncertainty of the
% \red{QT based}  metrology estimation for our weak target field is then
% $\delta \omega _{n,t}\simeq {e^{{\gamma^2 t^2}/{n}}}/ { \sqrt{TtL}
% (1-p)^{n-1}}$.
\textcolor{black}{For a short time such as  $T\ll  {1}/{\sqrt{p}\gamma }$,
the error due to QT is negligible, and so we obtain the same results as
with perfect QT by setting $t=T$and $n=4\gamma ^2T^2$,}
% $\delta \omega _{T,{1}/{4p}}\simeq   {e^{{1}/{4}+{\gamma
% ^2 T^2}/{n}}}/{T\sqrt{L}}\leq {e^{{1}/{2}}}/{T\sqrt{L}}$ when
% $t=T$ and $n={1}/{4p}$.
 which allows us to achieve 
uncertainty scaling as $1/T$.
% \textcolor{blue}{We can rewrite the condition for the error to be independent ($\tilde{n}\tau
%  \geq \tau _c$) as $pLT\gg \tau _c$, which can be satisfied for a large $L$.} 
%Heisenberg limited sensitivity. 
% In quantum metrology, another regime that is quite often
% considered is the scaling law in the limit of long T (much greater than the coherence time of the system). We will consider this now.
%\red{In quantum metrology} one typically considers the  scaling law in
%the limit of long $T$, which we will now discuss.
We can minimize the uncertainty
% \textcolor{black}{$\delta \omega _{n,t}\simeq {e^{{\gamma^2 t^2}/{n}}}/ { \sqrt{TtL}
% (1-p)^{n-1}}$}
by setting
$t_{\rm{opt}}={\sqrt{n}}/{2\gamma}$ as long as $T\gg t_{\rm{opt}}$ is
satisfied.
\red{In such a case,}  $\delta
\omega_{n,t_{\rm{opt}}}=\frac{\sqrt{2}e^{{1}/{4}}}{(1-p)^{n-1}}
{\sqrt{\frac{\gamma }{\sqrt{n} TL}}}$,
  which for $n=1$ gives the standard Ramsey uncertainty $\delta
 \omega_{{\rm{R}}}=\frac{e^{{1}/{4}}\sqrt{\gamma}}{\sqrt{TL}}$
 \cite{huelga1997improvement}
 where we replace $L$ with $2L$ (because the standard Ramsey scheme can
 utilize every qubit  to probe the target fields without ancillary
 qubits).
 For $n\gg 1$.  we can treat $n$ as a continuous variable, and we can
 analytically minimize the uncertainty as $\delta
 \omega_{\rm{opt}}\simeq 2\sqrt{\frac{e\sqrt{p }\gamma }{LT}}$ for
${1}/{16\gamma ^2T^2}\ll p\ll 1$ where we choose
 $n_{\rm{opt}}=-{1}/{4\log (1-p)}\simeq {1}/{4p}$. \textcolor{black}{The
 condition required for the error to be independent
 ($\tilde{n}\tau
 \geq \tau _c$) is written as $L\sqrt{p}/\gamma \gg \tau _c$, and is
 satisfied for a large $L$.}
% is satisfied just by increasing the number of the
% qubits where we have $\tilde{n}=2L-1$ and $\tau =t_{\rm{opt}}/n_{\rm{opt}}$. 
 In this case, we have a constant factor improvement over the standard
 Ramsey scheme for a longer $T$. In fact, as long as $p<0.0251$,
 our scheme is better than that standard Ramsey scheme (${\delta
 \omega_{{\rm{R}}}}/{\delta \omega_{{\rm{opt}}}}>1$).
 %Also,
 %If we have
 For $p=10^{-4}$, we obtain ${\delta
 \omega_{{\rm{R}}}}/{\delta \omega_{{\rm{opt}}}}\simeq 3.89$,
 %\textcolor{black}{and so our sensor has a clear advantage}.
 So our sensor has an advantage with  finite errors caused by the
 imperfect QT.

In conclusion, we have proposed
%\red{in this letter}
a scheme designed to achieve
%Heisenberg limited
\textcolor{black}{sensitivity beyond the classical limit and to measure}
%quantum sensing of
the amplitudes of globally
applied fields.
We have found that frequent implementations of quantum teleportation
provide a suitable circumstance for sensing where the dephasing is
suppressed while the information from the target fields is continuously
accumulated.
If perfect quantum teleportation is available,
\textcolor{black}{the uncertainty scales as $1/T$ with our scheme while any classical
sensor shows the uncertainty scaled as $1/\sqrt{T}$}.
%we can
%achieve the
%Heisenberg limit.
Moreover, even when quantum teleportation is moderately noisy, our
protocol still realizes superior quantum enhancement to the standard Ramsey scheme.
% \begin{table}[htb]
%  \centering
%  \begin{tabular}{|l|c|r|r|}\hline
%     &{\rm{perfect QT}}&imperfect QT for a short T&imperfect QT for a long T \\ \hline \hline
%     separable&1&1&1 \\ \cline{2-4} 
%     entanglement&1&1&1 \\ \hline
% \end{tabular}
%  \end{table}
This work was supported by JSPS KAKENHI Grant No.
15K17732 and partly supported by MEXT KAKENHI Grant
No. 15H05870. S.C.B. acknowledges support from the EPSRC
NQIT Hub, Project No. EP/M013243/1.

\newpage
\appendix{\bf{Supplementary materials:}}
This is a supplementary material of the paper titled
 ``Quantum metrology beyond the classical limit
under the effect of dephasing''.
\textcolor{black}{
\section{Error model}
In the main text, we consider a specific noise model with which to calculate the
sensitivity of our sensor, and we will explain how we can derive the
model from the general expression.
We consider the following general Hamiltonian to describe the
dephasing.
\begin{eqnarray}
 H&=&H_{\rm{S}}+H_{\rm{I}}+H_{\rm{E}}\nonumber \\
 H_{\rm{S}}&=&\frac{\omega }{2}\sigma _z\otimes
  \openone_{\rm{E}} \nonumber \\
 H_{\rm{E}}&=&\openone_{\rm{S}} \otimes C
  \nonumber \\
\end{eqnarray}
 $H_{\rm{S}}$, $H_{\rm{I}}$, and $H_{\rm{E}}$
 denote a system, an interaction, and an environmental Hamiltonian,
 respectively.
 As we describe in the main text, the perturbation theory allows us to solve
 the Schrodinger equation, and we obtain
 \begin{eqnarray}
  \rho _{\rm{S}}(\tau )\simeq (1-\epsilon _\tau )U_{\tau
}\rho_{\rm{S}}(0)U_{\tau }^{\dagger }+\epsilon _\tau \hat{\sigma
}_z  U_{\tau }\rho _{\rm{S}}(0) U_{\tau}^{\dagger
}\sigma _z \label{general}
 \end{eqnarray}
 where $\epsilon _{\tau }=\lambda ^2 C_0 \tau ^2$
denotes an error rate for $\lambda ^2 C_0 (\tau)^2 \ll 1$ and
$U_{\tau}=e^{-{i\omega \tau}\hat{\sigma }_z/2}$
denotes a unitary operator. We could approximate this expression by
\begin{eqnarray}
 \rho _{\rm{S}}(\tau ) \simeq \frac{1+e^{-2\lambda ^2 C_0 \tau^2}}{2}U_{\tau
}\rho_{\rm{S}}(0)U_{\tau}^{\dagger}\nonumber \\
+\frac{1-e^{-2\lambda ^2 C_0 \tau^2}}{2} \hat{\sigma }_z
U_{\tau}\rho_{\rm{S}}(0)U_{\tau}^{\dagger} \sigma _z \label{specific}
\end{eqnarray}
Actually, by performing a Taylor expansion such as $e^{-2\lambda ^2 C_0
\tau^2}\simeq 1-2\lambda ^2 C_0 \tau^2$ in Eq. (\ref{specific}), we
can reproduce Eq. (\ref{general}).
By defining
$\gamma^2
=2\lambda ^2 C_0$, we obtain
$\rho _{\rm{S}}(\tau ) \simeq \frac{1+e^{-\gamma ^2 \tau^2}}{2}U_{\tau
}\rho_{\rm{S}}(0)U_{\tau}^{\dagger}
+\frac{1-e^{-\gamma ^2 \tau^2}}{2} \hat{\sigma }_z
U_{\tau}\rho_{\rm{S}}(0)U_{\tau}^{\dagger} \sigma _z $.
In the main text, we use this expression to quantify the effect of the
dephasing during the operations in our sensing protocol.
It is worth mentioning that, strictly speaking, the form of the
Eq. \ref{specific} is the same as
Eq. \ref{general} only when $\gamma \tau \ll 1$.
However, typical dephasing models
\cite{PSE,de2010universal,KakuyanagiMenoSaitoNakanoSembaTakayanagiDeppeShnirman01a,YoshiharaHarrabiNiskanenNakamura01a}
actually show the behavior described in Eq. (\ref{specific}) 
for an even longer time
if the correlation time of the environment is much
longer than the dephasing time, and this shows the validity of our assumption.}

\section{Independent error accumulation}
In our scheme, the environment interacts with
\textcolor{black}{teleported}
states of the
qubit
one after another.
Here, we consider the decoherence dynamics for this case in more detail
than in
the main text. Specifically, we will
show that, although the environment frequently interacts with the
teleported states, the phase error on the qubit can be independent as
long as a Born approximation is valid. 

For simplicity, we consider just two sites for a Hamiltonian of the
form $H=H_{\rm{S}}+H_{\rm{I}}+H_{\rm{E}}$
 where
 $H_{\rm{S}}=\sum_{j=1}^{2} \frac{\omega }{2}\sigma ^{(j)}_z $
 ($H_{\rm{E}}=\sum_{j=1}^{2}\openone^{(j)}_{\rm{S}} \otimes C_j$) 
 denotes \red{the} system
(environmental) Hamiltonian,
$H_{\rm{I}}=\sum_{j=1}^{2}\lambda \sigma ^{(j)}_z\otimes B_j$
 denotes an
interaction between the system and
the environment. 
$B_j$ and $C_j$ denote the environmental operators.
In the interaction picture, we have
$ H_{\rm{I}}(t)=\lambda \sum_{j=1}^{2}\sigma ^{(j)}_z\otimes
 \tilde{B}_j(t)$
 where
 $\tilde{B}_j(t)=e^{iH_{\rm{E}}t}B_je^{-iH_{\rm{E}}t}$.
 The initial state is given as $\rho (0)=  \rho
 _{\rm{S}}(0)\otimes \rho
 _{\rm{S}}(0) \otimes \rho^{(1)}_{\rm{E}}\otimes \rho^{(2)}_{\rm{E}}$.
  We also assume that $\rho^{(j)}_{\rm{E}}$ is in thermal
 equilibrium such that $[\rho^{(j)}_{\rm{E}},H_{\rm{E}}]=0$ and
 our noise  is non-biased such that
 ${\rm{Tr}}[\rho^{(j)}_{\rm{E}}B_j] =0$.
 %We consider a case that
 We consider the following case.
 After the environment interacts with the system qubit at each site  for a
 time $0\leq t \leq \tau $, we perform a
 quantum teleportation  to send the state at site 1 to site 2
 (while the state at site 2 is
 \textcolor{black}{teleported}
 to another site), and
 the environment interacts with the new state at each site for a time
 $\tau \leq t \leq 2 \tau  $.
 To describe the interaction between the system
 and environment before the quantum teleportation, we use the Schrodinger equation 
 \begin{eqnarray}
  \frac{d\rho _I(t)}{dt}=-i[H_I(t),\rho _I(t)].
 \end{eqnarray}
 By integrating this, we obtain
 \begin{eqnarray}
  \rho _I(t)&=&\rho (0)-i\int_{0}^{t}dt'[H_I(t'), \rho (0)]\nonumber \\
   &-&i\int_{0}^{t}\int_{0}^{t'}dt'dt''[H_I(t'), [H_I(t''), \rho
    (t'')]]\nonumber \\
 \end{eqnarray}
 Since there are no interactions between site 1 and site 2, we
 consider that the state at site 1 is separable from the state at
 the site 2.
 \begin{eqnarray}
 \rho ^{(j)}_{\rm{I}}(\tau )\simeq \rho ^{(j)}(0)-i\lambda
  \int_{0}^{\tau }dt'\lambda [\sigma ^{(j)}_z\otimes
  \tilde{B}_j(t'),\rho (0)]\nonumber \\
 -\lambda ^2\int_{0}^{\tau }\int_{0}^{t'}dt'dt'' [\sigma ^{(j)}_z\otimes
  \tilde{B}_j(t'),[\sigma ^{(j)}_z\otimes
  \tilde{B}_j(t''),\rho (0)]]\nonumber
\end{eqnarray}
for $j=1,2$
where we use a second order perturbation expansion in $\lambda $.
We use the Born approximation $\rho ^{(j)}_{\rm{I}}(\tau ) \simeq \rho
^{(j)}_{\rm{S}}(\tau ) \otimes \rho^{(j)}_{\rm{E}} $
\cite{hornberger2009introduction}. This means that, since the environment
has a large degree of freedom, the correlation between the system and
the environment is negligible. In fact, it is known that
%, as we increase the size of the bath,
the Born approximation becomes more accurate as the size of the bath is increased \cite{hasegawa2011classical,carcaterra2011dissipation}.
By tracing out the environment, we obtain
\begin{eqnarray}
\rho ^{(j)}_{\rm{S}}(\tau )&\simeq& \rho _{\rm{S}}(0) \nonumber \\
&& -\lambda ^2\int_{0}^{\tau }\int_{0}^{t'}dt'dt''  C^{(j)}_{t'-t''} \left[\hat{\sigma }^{(j)}_z,\left[\hat{\sigma }^{(j)}_z,\rho _{\rm{S}}(0)\right]\right]
  \nonumber
\end{eqnarray}
for $j=1,2$
where $C^{(j)}_{t'-t''}\equiv
\frac{1}{2}{\rm{Tr}}[(\tilde{B}_j(t')\tilde{B}_j(t'')+\tilde{B}_j(t'')\tilde{B}_j(t'))
\rho^{(j)}_{\rm{E}}]$
denotes the correlation function of the environment at the site $j$. It
is worth mentioning that the correlation function does not depend on the
state of the system but depends on the properties of the environment.
By performing a
 quantum teleportation to send the state at site 1 to site 2
 (while the state at site 2 is
  \textcolor{black}{teleported}
 to another site), we obtain
\begin{eqnarray}
\rho '^{(2)}_{\rm{S}}(\tau )&\simeq& \rho _{\rm{S}}(0) \nonumber \\
&& -\lambda ^2\int_{0}^{\tau }\int_{0}^{t'}dt'dt''  C^{(1)}_{t'-t''}
 \left[\hat{\sigma }^{(2)}_z,\left[\hat{\sigma }^{(2)}_z,\rho _{\rm{S}}(0)\right]\right]
  \nonumber
\end{eqnarray}
at site 2.
By allowing this state to interact with the environment for a time $\tau \leq t \leq
2 \tau $, we obtain
 \begin{eqnarray}
 \rho '^{(2)}_{\rm{I}}(2\tau )\simeq \rho _{\rm{I}}'^{(2)}(\tau )-i\lambda
  \int_{\tau }^{2\tau }dt'\lambda [\sigma ^{(2)}_z\otimes
  \tilde{B}_2(t'),\rho '^{(2)}_{\rm{I}}(\tau )]\nonumber \\
 -\lambda ^2\int_{\tau }^{2\tau }\int_{\tau }^{t'}dt'dt'' [\sigma ^{(2)}_z\otimes
  \tilde{B}_2(t'),[\sigma ^{(2)}_z\otimes
  \tilde{B}_2(t''),\rho _{\rm{I}}'^{(2)}(\tau )]]\nonumber
\end{eqnarray}
where we consider $\rho _{\rm{I}}'^{(2)}(\tau )=\rho '^{(2)}_{\rm{S}}(\tau )
\otimes \rho _{\rm{E}}^{(2)}$ as the initial state.
By tracing out the environment, we obtain
\begin{eqnarray}
&&\rho '^{(2)}_{\rm{S}}(2\tau )\simeq \rho '^{(2)}_{\rm{S}}(\tau ) \nonumber \\
 &&-\lambda ^2\int_{\tau }^{2\tau }\int_{\tau}^{t'}dt'dt''  C^{(2)}_{t'-t''}
 \left[\hat{\sigma }^{(2)}_z,\left[\hat{\sigma }^{(2)}_z,\rho'^{(2)}_{\rm{S}}(\tau )\right]\right]
  \nonumber 
\end{eqnarray}
where we use a Born approximation such as $\rho _{\rm{I}}'^{(2)}(2\tau )\simeq \rho '^{(2)}_{\rm{S}}(2\tau )
\otimes \rho _{\rm{E}}^{(2)}$.
By  considering up to the second order term of $\lambda$, we
obtain
\begin{eqnarray}
&&\rho '^{(2)}_{\rm{S}}(2\tau )\simeq \rho _{\rm{S}}(0 )
 \nonumber \\
  &-&\lambda ^2\int_{0}^{\tau }\int_{0}^{t'}dt'dt''  C^{(1)}_{t'-t''}
 \left[\hat{\sigma }^{(2)}_z,\left[\hat{\sigma }^{(2)}_z,\rho _{\rm{S}}(0)\right]\right]
 \nonumber \\
 &-&\lambda ^2\int_{\tau }^{2\tau }\int_{\tau}^{t'}dt'dt''  C^{(2)}_{t'-t''}
 \left[\hat{\sigma }^{(2)}_z,\left[\hat{\sigma }^{(2)}_z,\rho
                              _{\rm{S}}(0 )\right]\right].
  \nonumber 
\end{eqnarray}
Since $C^{(2)}_{t'-t''}$ does not depend on the absolute value of $t'$
(or $t''$) but depends on the time difference $t'-t''$, we obtain
\begin{eqnarray}
&&\rho '^{(2)}_{\rm{S}}(2\tau )\simeq \rho _{\rm{S}}(0 )
 \nonumber \\
  &-&\lambda ^2\int_{0}^{\tau }\int_{0}^{t'}dt'dt''  C^{(1)}_{t'-t''}
 \left[\hat{\sigma }^{(2)}_z,\left[\hat{\sigma }^{(2)}_z,\rho _{\rm{S}}(0)\right]\right]
 \nonumber \\
 &-&\lambda ^2\int_{0 }^{\tau }\int_{0}^{t'}dt'dt''  C^{(2)}_{t'-t''}
 \left[\hat{\sigma }^{(2)}_z,\left[\hat{\sigma }^{(2)}_z,\rho
                              _{\rm{S}}(0 )\right]\right].
  \nonumber 
\end{eqnarray}
So the interaction between the system and the environment at
site 1 for a time $0\leq t\leq \tau $ does not induce a correlation in the
decoherence dynamics between the system and environment at
site 2. (There is no effect such as an accelerated or deaccelerated
noise accumulation of the dephasing at site 2 due to past
decoherence at site 1.)
So the $\hat{\sigma }_z$ error caused by the environment
at site $2$ acts on the qubit independently of the $\hat{\sigma }_z$
error caused by the environment
at site $1$.

Although we consider the dynamics at sites $1$ and $2$ above, we can easily
generalize this to $n$ sites.
By repeating the above calculations, we obtain
\begin{eqnarray}
&&\rho '^{(n)}_{\rm{S}}(n\tau )\simeq \rho _{\rm{S}}(0 )
 \nonumber \\
  &-&\sum_{j=1}^{n}\lambda ^2\int_{0}^{\tau }\int_{0}^{t'}dt'dt''  C^{(j)}_{t'-t''}
 \left[\hat{\sigma }^{(n)}_z,\left[\hat{\sigma }^{(n)}_z,\rho _{\rm{S}}(0)\right]\right]
 \nonumber \\
\end{eqnarray}
at site $n$ after $n-1$ teleportation where the $\hat{\sigma
}_z$ error accumulates independently at each site, and this is
consistent with the calculations in the main text.

\section{Effect of a finite number of the qubits}
In the main text,
%we consider a limit of a large number of the qubits,
%which is a common assumption in the field of quantum metrology.
%In this case,
%In the main text,
%we consider a case of
%$n\ll 2L$, and so
we did not mention the case where
the qubits are
\textcolor{black}{teleported}
back to the original sites whose environment
the qubit initially interacted with. 
%However,
If
%we consider a smaller
%number of the qubits,
%we have a condition of $n> 2L$, 
qubits have the chance to interact with the environment with which they previously interacted,
%with the same environment as that with which the qubits interacted
%before,
this may induce a correlated error on the qubit.
However, we will show that we can still avoid the
correlated error as long as there are a large number of the qubits, which is a common assumption in the field of quantum metrology..
%We consider
%a condition when we can avoid such a correlated error.

For simplicity, we consider just a single site for a Hamiltonian of the
form $H=H_{\rm{S}}+H_{\rm{I}}+H_{\rm{E}}$
 where
 $H_{\rm{S}}= \frac{\omega }{2}\sigma _z $
 ($H_{\rm{E}}=\openone_{\rm{S}} \otimes C_j$) 
 denotes \red{the} system
(environmental) Hamiltonian, and
$H_{\rm{I}}=\lambda f(t)\sigma _z\otimes B_j$
 denotes an
interaction between the system and the
environment. 
$B$ and $C$ denote environmental operators.
We define  $f(t)$ as follows

\[
  f(t) = \begin{cases}
    1 & (0\leq t \leq \tau  ) \\
              1 & ( (m-1)\tau \leq t \leq m\tau ) \\
    0 & ({\rm{otherwise}})
  \end{cases}
\] where $m$ denotes a natural number. This means that, after the qubit interacts with the environment for
    a time $(0\leq t \leq \tau)$, the qubit is decoupled from the
    environment for a time
    $\tau < t<n\tau $, and the qubit interacts with the same environment
    again for a time $m\tau \leq t \leq (m+1)\tau$. From this
    calculation, we can estimate the way in which the correlated error will be
    induced by multiple interactions with the same environment in our
    scheme. Using a similar calculation to that used in the previous section, we obtain
\begin{eqnarray}
&&\rho _{\rm{S}}(n\tau  )\simeq \rho _{\rm{S}}(0) \nonumber \\
&& -\lambda ^2\int_{0}^{m\tau  }\int_{0}^{t'}dt'dt''  f(t')f(t'')C_{t'-t''} \left[\hat{\sigma }_z,\left[\hat{\sigma }_z,\rho _{\rm{S}}(0)\right]\right]
  \nonumber
\end{eqnarray}
where we solve the Schrodinger equation for a time $0\leq t \leq m\tau
$. We obtain
\begin{eqnarray}
&&\rho _{\rm{S}}(m\tau  )-\rho _{\rm{S}}(0)\simeq  \nonumber \\
&& -\lambda ^2( \int_{0}^{\tau  } +\int_{(m-1)\tau }^{m\tau  })dt'\int_{0}^{t'}dt''  f(t')f(t'')C_{t'-t''} \left[\hat{\sigma }_z,\left[\hat{\sigma }_z,\rho _{\rm{S}}(0)\right]\right]
  \nonumber \\
&&= -\lambda ^2\int_{0}^{\tau  } dt'\int_{0}^{t'}dt''  f(t')f(t'')C_{t'-t''} \left[\hat{\sigma }_z,\left[\hat{\sigma }_z,\rho _{\rm{S}}(0)\right]\right]
  \nonumber \\
 && -\lambda ^2\int_{(m-1)\tau }^{m\tau  } dt'\int_{0}^{t'}dt''  f(t')f(t'')C_{t'-t''} \left[\hat{\sigma }_z,\left[\hat{\sigma }_z,\rho _{\rm{S}}(0)\right]\right]
  \nonumber \\
 &&= -\lambda ^2\int_{0}^{\tau  } dt'\int_{0}^{t'}dt''  C_{t'-t''} \left[\hat{\sigma }_z,\left[\hat{\sigma }_z,\rho _{\rm{S}}(0)\right]\right]
  \nonumber \\
 && -\lambda ^2\int_{(m-1)\tau }^{m\tau  } dt'\int_{0}^{\tau }dt''  C_{t'-t''} \left[\hat{\sigma }_z,\left[\hat{\sigma }_z,\rho _{\rm{S}}(0)\right]\right]
  \nonumber \\
  && -\lambda ^2\int_{(m-1)\tau }^{m\tau  } dt'\int_{(n-1)\tau }^{t' }dt''  C_{t'-t''} \left[\hat{\sigma }_z,\left[\hat{\sigma }_z,\rho _{\rm{S}}(0)\right]\right]
  \nonumber 
\end{eqnarray}
The term $\lambda ^2\int_{(m-1)\tau }^{m\tau  } dt'\int_{0}^{\tau }dt''
C_{t'-t''} \left[\hat{\sigma }_z,\left[\hat{\sigma }_z,\rho
_{\rm{S}}(0)\right]\right]$ represents the correlated effect of an
environment that has a memory of the first interaction with the qubit
when it has the second interaction with the qubit.
It is worth mentioning that a typical environment has a finite
correlation time $\tau _c$
\cite{romach2015spectroscopy,de2010universal,kondo2016using,wilhelm2007superconducting,hall2014analytic,koppens2007universal},
and so the correlation function for a
longer time scale than the correlation time such as $C_{t'-t''} \simeq
0$ for $|t'-t''|\gg \tau _c$ has a negligible effect.
This means that, if we have $m\tau \gg \tau _c$, we obtain
\begin{eqnarray}
&&\rho _{\rm{S}}(m\tau  )\simeq  \rho _{\rm{S}}(0)\nonumber \\
 && -\lambda ^2\int_{0}^{\tau  } dt'\int_{0}^{t'}dt''  C_{t'-t''} \left[\hat{\sigma }_z,\left[\hat{\sigma }_z,\rho _{\rm{S}}(0)\right]\right]
  \nonumber \\
  && -\lambda ^2\int_{(m-1)\tau }^{m\tau  } dt'\int_{(m-1)\tau }^{t' }dt''  C_{t'-t''} \left[\hat{\sigma }_z,\left[\hat{\sigma }_z,\rho _{\rm{S}}(0)\right]\right]
  \nonumber \\
 &\simeq&  \rho _{\rm{S}}(0) -2\lambda ^2\int_{0}^{\tau  } dt'\int_{0}^{t'}dt''  C_{t'-t''} \left[\hat{\sigma }_z,\left[\hat{\sigma }_z,\rho _{\rm{S}}(0)\right]\right],
  \nonumber 
\end{eqnarray}
and the correlated effect of the environment becomes negligible.
%Therefore,
So the decoherence acts on the qubit independently as long as we
have $m\tau \gg \tau _c$.
\textcolor{black}{We define $\tilde{n}$
as
the maximum teleportation number without
%the maximum number of the teleportation without
the qubit state
being teleported back to the original site.
(For example, in the teleportation-based scheme with a ring
structure described in the main text for separable states, we have
$\tilde{n}=2L-1$.)
We can substitute $m=\tilde{n}$, and the
condition needed for the error to be independent is $\tilde{n} \tau \gg \tau _c$.}
%Therefore, as long as we have a large number of the qubits, we can
%avoid the correlated error.
% \textcolor{blue}{On the other hand, for the
% teleportation-based scheme for entangles states, we set
% $m=\frac{L}{M}$. So, to consider the error as independent, we need a condition of
%   $(L/M)\tau \gg \tau _c \Leftrightarrow 1\gg c^2L\gamma ^2
%   T\tau _c$ where we choose $t=T$, $M=cL$, and $n=M\gamma^2T^2$, which can be satisfied only for a small size of the
%   entangled state.}
%It is worth mentioning that, since we can set $t=T$ if perfect quantum
%teleportation is .
% Since we
% need a condition of $\frac{\gamma ^2T^2}{n}\ll 1$ to achieve the
% Heisenberg limit, the necessary number of the qubits is given as
% $\gamma ^2T^2 n\tau = \gamma ^2T^3 \gg  \tau _c$.

\section{High frequency noise}
\begin{figure}[h!] 
\begin{center}
\includegraphics[width=0.8\linewidth]{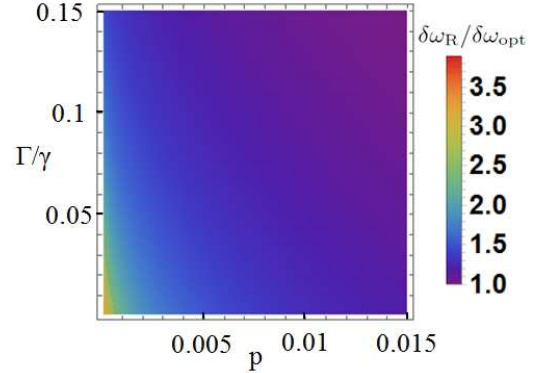} 
\caption{
Performance of our teleportation based scheme against the standard
 Ramsey scheme. We plot $\delta \omega _{\rm{R}}/\delta
 \omega_{\rm{opt}}$ against $\frac{\Gamma }{\gamma }$ and $p$. Our
 scheme performs better than the standard Ramsey scheme for
 \textcolor{black}{$\Gamma /\gamma <0.130$ and $p<0.0123$ }.
 %\red{MAYBE IN THE FIGURE PLOT ONLY  $P<0.01$ and $\Gamma<0.01$}
 }
 \label{tone}
\end{center}
\end{figure}

In the main text, we consider the effect of dephasing as decoherence.
There are of course other sources of decoherence that cannot be
suppressed by the quantum teleportation (QT) protocol, and we consider such an effect.
%\red{For instance},
If our quantum systems are affected by
high-frequency noise with a short correlation time,
the decay is not quadratic in \red{nature but more exponential-like}. Energy relaxation in a high temperature environment
is known to induce such noise \cite{GZ01b}. \red{Consider the situation
in which} an initial state $|+\rangle _{j}$ evolves
under the effect of both low-frequency dephasing and high-frequency
noise for a time $\tau $.
In such a case
$\rho ^{(j)}_{\rm{S}}(\tau) =\frac{1}{2}(\openone _{j}+|1\rangle
_{j}\langle0|e^{-i\omega \tau  -\gamma ^2\tau ^2-\Gamma \tau}
+|0\rangle _{j}\langle 1|e^{i\omega \tau -\gamma ^2\tau^2-\Gamma \tau})$
where $\Gamma $ denotes the decay rate \red{associated}
with the high-frequency noise ($\Gamma =0$ gives the same noise model as
the one we
used previously). \red{It is then straightforward to}
calculate the uncertainty of the estimation under the effect of this
noise with imperfect QT as
$\delta \omega _{n,t}=\frac{e^{\Gamma t +{\gamma ^2t^2}/{n}}}{(1-p)^{n-1}}\frac{1}{\sqrt{TtL}}$.
Choosing $t_{\rm{opt}}=\frac{-\sqrt{n}m+\sqrt{nm^2+4}}{4(\gamma /\sqrt{n})}$
(where $m={\Gamma }/{\gamma}$), we minimize this \red{with respect to} time as
\begin{eqnarray}
 \delta \omega_{n,t_{\rm{opt}}}=\frac{2\sqrt{\gamma}e^{\frac{1}{4}+\frac{-nm^2+m\sqrt{n(nm^2+4)}}{8}}}{(1-p)^{n-1}\sqrt{(-nm+\sqrt{n(nm^2+4)})LT}}.\nonumber
\end{eqnarray}
The uncertainty with $n$ can be numerically minimized
%A numerical minimization of the uncertainty with $n$
as $\delta
\omega_{\rm{opt}}=\min _n \delta \omega_{n,t_{\rm{opt}}}$.
%\textcolor{black}{can be done}.
 \red{
 %and so
 \textcolor{black}{In}
Fig. \ref{tone}, we plot $\delta \omega _{\rm{R}}/\delta
\omega_{\rm{opt}}$ versus $p$ and ${\Gamma }/{\gamma}$ where $\delta
\omega _{\rm{R}}$ is the uncertainty for the standard Ramsey scheme. Our
plots shows} that our scheme performs better than the standard
Ramsey scheme for
\textcolor{black}{$\Gamma /\gamma <0.130$ and $p<0.0123$ }.

\section{Implementation of quantum metrology beyond the classical limit under
 the effect of dephasing with global control}
In the scheme described in the main text, we suggest the use of  quantum
teleportation to suppress dephasing.
%This clearly illustrates how qubit
%motion can enhance the sensitivity of the metrology.
%However,
To implement this scheme, the
quantum teleportation. which involves projective measurements and
feedback operations,
should be performed in a much shorter time than the correlation
time of the environment. Also, the individual controllability of the
qubit is needed to implement quantum teleportation for every
qubit.
% Such a rapid and accurate control of the qubit
% might be difficult to realize in the
% current technology.
% However,
 In the state of the art technology, a gate
control with a fidelity of more than 99\% has been realized for some
systems such as superconducting qubits and ion trap systems
\cite{barends2014superconducting,ballance2016high,blume2017demonstration}.
Moreover, many groups aim to realize a scalable quantum computer, and a
quantum supremacy with 50 qubits could be demonstrated in the near
future \cite{neill2017blueprint,popkin2016quest}. Such a development of
quantum technology toward the demonstration of quantum computation
also supports the realization of our teleportation-based scheme.
So our scheme is
within the reach of such a future technology.

However, it is still worth discussing how to realize our scheme with
simpler technology and thus improve the practicality.
In this section, we suggest an alternative
scheme for this purpose, which is useful for our separable states scheme.
%to improve the practicality.
Here, we use a direct interaction between the qubits. We
show that a
simple modulation of the Hamiltonian can transfer the states to suppress
the dephasing where neither individual addressability nor
rapid measurement is required. 
% let the qubit state travel  
% This might be difficult to realize in the
% current technology, because such a scheme requires rapid measurements
% and feedback on every qubit where individual control .
\begin{figure}[h!] 
\begin{center}
\includegraphics[width=0.99\linewidth]{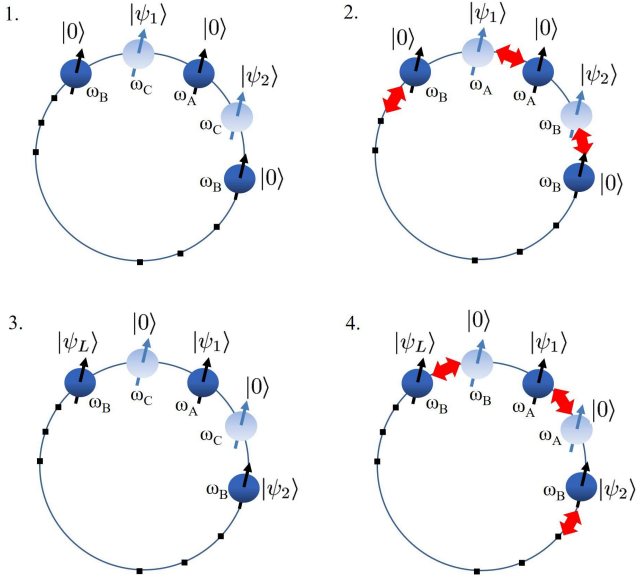} 
\caption{Schematic illustration of our SWAP-based sensing where we arrange $2L$ qubits in a ring structure.
   Similar to the teleportation-based scheme described in the main text, half of the qubits contain an information of the target fields as a
 probe while the remaining half are used as an ancilla qubits kept in a
 ground state.
 We
 assume a flip-flop type interaction between nearest neighbor qubits. For the half of the qubits, the frequency is fixed at
 $\omega _A$ or $\omega _B$ where $\omega _A$ is well detuned from
 $\omega _B$, while a frequency of the
 other half of the qubits is tunable. If we set the tunable frequency as $\omega
 _A$ or $\omega _B$, the flip-flop interaction becomes effective, and so
 we can perform a SWAP gate between the probe qubits and ancillary
 qubits by a time evolution with the Hamiltonian. On the
 other hand, we set the tunable frequency as $\omega _C$ where $\omega
 _C$ is well detuned from both $\omega _A$ and $\omega _B$, the
 flip-flop interaction is significantly suppressed by the detuning. In
 this configuration, the modulation of the
 frequency can transfer the state of the probe qubits
 to the clock wise direction.
 }
 \label{swapfigure}
\end{center}
\end{figure}

%Our basic idea is to use a SWAP gate between the qubits.
The main difference from the scheme described in the main text
is that, instead of quantum teleportation, we use a SWAP operation
between the probe qubit at a site and an ancillary qubit
at the 
 neighboring site on the right,
as shown in Fig. \ref{swapfigure}.
First,  we arrange qubits in a ring
structure, and
we prepare a state of $\bigotimes _{j=1}^L|+\rangle
_{2j-1}$  located at site $2j-1$ ($j=1,2,\cdots ,L$) for probe qubits, while we
prepare $\bigotimes _{j=1}^L|0\rangle
_{2j}$  located at  site $2j$ ($j=1,2,\cdots ,L$) for the ancillary
qubits. For simplicity, we assume that $L$ is an even number.
% we prepare a probe state of $\bigotimes _{j=1}^L|+\rangle
% _{2j-1}$  located at the site $2j-1$ ($j=1,2,\cdots ,L$).
 Second, we
then let the  state evolve for a time $\tau ={t}/{n}$ and then perform a
SWAP gate
between the probe qubit and the
ancillary qubit at the
 neighboring site on the right (we assume that our gate operations are much faster than
$\tau$). Third, we repeat the second step $(n-1)$
times while in the fourth step
we allow this state to evolve for a time $\tau =\frac{t}{n}$, and readout
the state by measuring $\hat{M}_y=\sum_{j=1}^{L}\hat{\sigma }_y^{(j)}$.
Finally, we repeat these steps $N$ times during the measurement time $T$
where $N\simeq {T}/{t}$ is the repetition number.

 \begin{figure}[h!] 
\begin{center}
\includegraphics[width=1.0\linewidth]{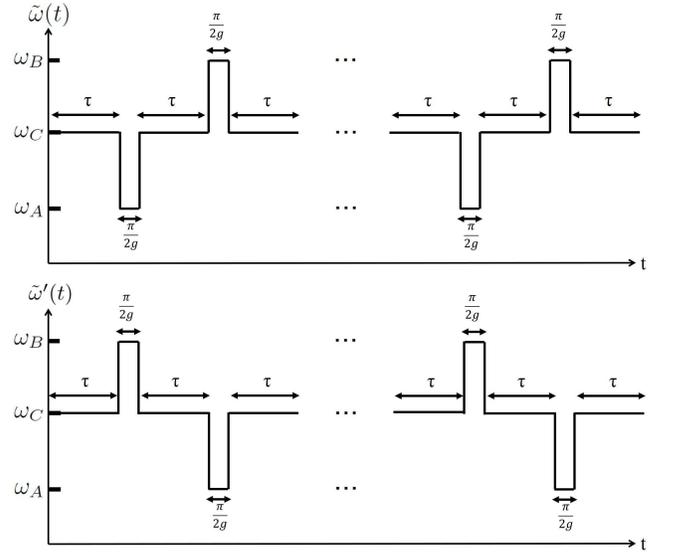} 
\caption{
Modulation of the frequency of the qubit to implement our SWAP-based
 scheme. The qubit frequency is either $\omega _A$, $\omega _B$, or
 $\omega _C$ where these frequencies are well detuned from each other.
 Also, $g$ denotes the coupling
 strength between the qubits and $\tau =\frac{t}{n}$ denotes an interaction
 time with the fields. Here, we assume $\tau \gg \frac{1}{g}$.
 The modulation of the frequency with a
 flip-flop type interaction provides us
 with a way to transfer the state of the probe qubit to the clock wise
 direction as shown in the Fig. \ref{swapfigure}.
 }
 \label{tune}
\end{center}
\end{figure}
Importantly, we
can implement the SWAP gate by using a direct interaction between the
qubits, and we can turn on/off the interaction by modulating the
frequency of the qubit. We consider the following flip-flop type Hamiltonian.
\begin{eqnarray}
 H=\sum_{j=1}^{2L}\frac{\omega +\omega _j}{2}\hat{\sigma
  }_z^{(j)}+g(\hat{\sigma }_+^{(j)}\hat{\sigma }_-^{(j+1)}+\hat{\sigma }_-^{(j)}\hat{\sigma }_+^{(j+1)})
\end{eqnarray}
where $\omega _j$ denotes the frequency of the qubit at the $j$ th
site. We also consider a periodic condition such as $\hat{\sigma }_{\pm
}^{(2L+1)}=\hat{\sigma }_{\pm }^{(1)}$. We set the frequency of the
qubit as
% \begin{eqnarray}
%  \tilde{\omega }_j&=&\omega _A \ \ \ \ \ {\rm{  for } \ j\equiv 2({\rm{mod}}
%   \ 4) }\nonumber \\
%   \tilde{\omega }_j&=&\omega _B \ \ \ \ \ {\rm{  for } \ j\equiv 0({\rm{mod}}
%   \ 4) }\nonumber \\
% \end{eqnarray}
 \[
  \omega _j = \begin{cases}
                           \tilde{\omega }(t) & j\equiv 1\ ({\rm{mod}}  \ 4)
  \\
    \omega _A &   j\equiv 2\ ({\rm{mod}}
  \ 4)  \\
    \tilde{\omega }'(t) & j\equiv 3\ ({\rm{mod}}  \ 4)
  \\
                         \omega _B                     &   j\equiv 4\ ({\rm{mod}}  \ 4)
  \end{cases}
\]
where $\tilde{\omega }(t)$ and $\tilde{\omega }'(t)$ denote a
tunable time-dependent frequency, as shown in Fig. \ref{tune}. In addition, we assume
a large detuning between $\omega _A$ and $\omega _B$.  
Here, we focus on two adjacent
qubits, and the Hamiltonian is given as
% \begin{eqnarray}
% H^{(k,k+1)}=\frac{\omega +\tilde{\omega }_k}{2}\hat{\sigma
%   }_z^{(k)}+\frac{\omega +\tilde{\omega }_{k+1}}{2}\hat{\sigma
%   }_z^{(k+1)}\nonumber \\
%   +g(\hat{\sigma }_+^{(k)}\hat{\sigma }_-^{(k+1)}+\hat{\sigma }_-^{(k)}\hat{\sigma }_+^{(k+1)})
% \end{eqnarray}
% We divide this Hamiltonian into two part
\begin{eqnarray}
 H^{(k,k+1)}&=&H^{(k,k+1)}_0+H^{(k,k+1)}_1\nonumber \\
  H^{(k,k+1)}_0&=&\frac{\omega +\tilde{\omega }_k}{2}\hat{\sigma
  }_z^{(k)}+\frac{\omega +\tilde{\omega }_{k+1}}{2}\hat{\sigma
  }_z^{(k+1)} \nonumber \\
 H^{(k,k+1)}_1&=&g(\hat{\sigma }_+^{(k)}\hat{\sigma }_-^{(k+1)}+\hat{\sigma }_-^{(k)}\hat{\sigma }_+^{(k+1)})
\end{eqnarray}
where one of the qubits is a probe and the other qubit is an ancilla.
In an interaction picture, we have
\begin{eqnarray}
 H^{(k,k+1)}_I=\frac{\tilde{\omega }_k-\tilde{\omega
  }_{k+1}}{2}\hat{\sigma }_z^{(1)}+g(\hat{\sigma }_+^{(k)}\hat{\sigma
  }_-^{(k+1)}+\hat{\sigma }_-^{(k)}\hat{\sigma }_+^{(k+1)})\nonumber 
\end{eqnarray}
If we have large detuning between these two qubits such as
$|\tilde{\omega }_k-\tilde{\omega }_{k+1}|\gg g$, we obtain
\begin{eqnarray}
 H_I^{(k,k+1)}\simeq \frac{\tilde{\omega }_k-\tilde{\omega }_{k+1}}{2}\hat{\sigma
  }_z^{(k)}
\end{eqnarray}
where the coupling is effectively turned off. On the other hand, if we
have a resonant condition such as $\tilde{\omega }_{k}=\tilde{\omega
}_{k+1}$
\begin{eqnarray}
 H_I^{(k,k+1)}= g(\hat{\sigma }_+^{(k)}\hat{\sigma }_-^{(k+1)}+\hat{\sigma }_-^{(k)}\hat{\sigma }_+^{(k+1)})
\end{eqnarray}
%in a rotating frame.
In this case, the interaction is turned on, and a
unitary operation $U=e^{iH^{(k,k+1)}\frac{\pi }{2g}}$ provides us with a SWAP gate
between the probe qubit and ancillary qubit up to local operations.
 This means that, if
 we set $\tilde{\omega} (t)=\omega _A$ ($\tilde{\omega }(t)=\omega _B$), a probe qubit and ancillary qubit
 with a frequency of $\omega _A$ ($\omega _B$) start the interaction
 under a resonant condition, while these qubits do not interact with the
 other qubits with a frequency of $\omega _B$ ($\omega _A$). On the
 other hand, if we set $\tilde{\omega }(t)=\omega _C$ where $|\omega
 _A-\omega _C|\gg g$ and $|\omega
 _B-\omega _C|\gg g$, the qubit does not interact with any other qubits due
 to the large detuning.
Therefore,  the simple modulation of the qubit frequency (described in
 Fig. \ref{tune})
realizes qubit
motion in the same manner as that described in the main text with 
quantum teleportation. This is much more practical than the
teleportation-based scheme, because neither individual addressability nor
a rapid measurement is required for the SWAP-based scheme.  
% Second we
% then let the  state evolve for a time $\tau ={t}/{n}$ and then swap the
% state of the probe qubit with the state of the ancillary qubit at the
% next site.

%\bibliography{6mylibrary.bib}

\begin{thebibliography}{48}
\expandafter\ifx\csname natexlab\endcsname\relax\def\natexlab#1{#1}\fi
\expandafter\ifx\csname bibnamefont\endcsname\relax
  \def\bibnamefont#1{#1}\fi
\expandafter\ifx\csname bibfnamefont\endcsname\relax
  \def\bibfnamefont#1{#1}\fi
\expandafter\ifx\csname citenamefont\endcsname\relax
  \def\citenamefont#1{#1}\fi
\expandafter\ifx\csname url\endcsname\relax
  \def\url#1{\texttt{#1}}\fi
\expandafter\ifx\csname urlprefix\endcsname\relax\def\urlprefix{URL }\fi
\providecommand{\bibinfo}[2]{#2}
\providecommand{\eprint}[2][]{\url{#2}}

\bibitem[{\citenamefont{Budker and Romalis}(2007)}]{budker2007optical}
\bibinfo{author}{\bibfnamefont{D.}~\bibnamefont{Budker}} \bibnamefont{and}
  \bibinfo{author}{\bibfnamefont{M.}~\bibnamefont{Romalis}},
  \bibinfo{journal}{Nature Physics} \textbf{\bibinfo{volume}{3}},
  \bibinfo{pages}{227} (\bibinfo{year}{2007}).

\bibitem[{\citenamefont{Balasubramanian and {\it{et
  al}}}(2008)}]{balasubramanian2008nanoscaleetal}
\bibinfo{author}{\bibfnamefont{G.}~\bibnamefont{Balasubramanian}}
  \bibnamefont{and} \bibinfo{author}{\bibnamefont{{\it{et al}}}},
  \bibinfo{journal}{Nature} \textbf{\bibinfo{volume}{455}},
  \bibinfo{pages}{648} (\bibinfo{year}{2008}).

\bibitem[{\citenamefont{Maze and {\it {et al}}}(2008)}]{maze2008nanoscaleetal}
\bibinfo{author}{\bibfnamefont{J.}~\bibnamefont{Maze}} \bibnamefont{and}
  \bibinfo{author}{\bibnamefont{{\it {et al}}}}, \bibinfo{journal}{Nature}
  \textbf{\bibinfo{volume}{455}}, \bibinfo{pages}{644} (\bibinfo{year}{2008}),
  ISSN \bibinfo{issn}{0028-0836}.

\bibitem[{\citenamefont{Degen et~al.}(2016)\citenamefont{Degen, Reinhard, and
  Cappellaro}}]{degen2016quantum}
\bibinfo{author}{\bibfnamefont{C.}~\bibnamefont{Degen}},
  \bibinfo{author}{\bibfnamefont{F.}~\bibnamefont{Reinhard}}, \bibnamefont{and}
  \bibinfo{author}{\bibfnamefont{P.}~\bibnamefont{Cappellaro}},
  \bibinfo{journal}{arXiv preprint arXiv:1611.02427}  (\bibinfo{year}{2016}).

\bibitem[{\citenamefont{Simon}(1999)}]{simon1999local}
\bibinfo{author}{\bibfnamefont{J.}~\bibnamefont{Simon}},
  \bibinfo{journal}{Advances in Physics} \textbf{\bibinfo{volume}{48}},
  \bibinfo{pages}{449} (\bibinfo{year}{1999}).

\bibitem[{\citenamefont{Chang et~al.}(1992)\citenamefont{Chang, Hallen,
  Harriott, Hess, Kao, Kwo, Miller, Wolfe, Van~der Ziel, and
  Chang}}]{chang1992scanning}
\bibinfo{author}{\bibfnamefont{A.}~\bibnamefont{Chang}},
  \bibinfo{author}{\bibfnamefont{H.}~\bibnamefont{Hallen}},
  \bibinfo{author}{\bibfnamefont{L.}~\bibnamefont{Harriott}},
  \bibinfo{author}{\bibfnamefont{H.}~\bibnamefont{Hess}},
  \bibinfo{author}{\bibfnamefont{H.}~\bibnamefont{Kao}},
  \bibinfo{author}{\bibfnamefont{J.}~\bibnamefont{Kwo}},
  \bibinfo{author}{\bibfnamefont{R.}~\bibnamefont{Miller}},
  \bibinfo{author}{\bibfnamefont{R.}~\bibnamefont{Wolfe}},
  \bibinfo{author}{\bibfnamefont{J.}~\bibnamefont{Van~der Ziel}},
  \bibnamefont{and} \bibinfo{author}{\bibfnamefont{T.}~\bibnamefont{Chang}},
  \bibinfo{journal}{Applied physics letters} \textbf{\bibinfo{volume}{61}},
  \bibinfo{pages}{1974} (\bibinfo{year}{1992}).

\bibitem[{\citenamefont{Poggio and Degen}(2010)}]{poggio2010force}
\bibinfo{author}{\bibfnamefont{M.}~\bibnamefont{Poggio}} \bibnamefont{and}
  \bibinfo{author}{\bibfnamefont{C.}~\bibnamefont{Degen}},
  \bibinfo{journal}{Nanotechnology} \textbf{\bibinfo{volume}{21}},
  \bibinfo{pages}{342001} (\bibinfo{year}{2010}).

\bibitem[{\citenamefont{Huelga et~al.}(1997)\citenamefont{Huelga, Macchiavello,
  Pellizzari, Ekert, Plenio, and Cirac}}]{huelga1997improvement}
\bibinfo{author}{\bibfnamefont{S.}~\bibnamefont{Huelga}},
  \bibinfo{author}{\bibfnamefont{C.}~\bibnamefont{Macchiavello}},
  \bibinfo{author}{\bibfnamefont{T.}~\bibnamefont{Pellizzari}},
  \bibinfo{author}{\bibfnamefont{A.}~\bibnamefont{Ekert}},
  \bibinfo{author}{\bibfnamefont{M.}~\bibnamefont{Plenio}}, \bibnamefont{and}
  \bibinfo{author}{\bibfnamefont{J.}~\bibnamefont{Cirac}},
  \bibinfo{journal}{Phys. Rev. Lett.} \textbf{\bibinfo{volume}{79}},
  \bibinfo{pages}{3865} (\bibinfo{year}{1997}).

\bibitem[{\citenamefont{Said et~al.}(2011)\citenamefont{Said, Berry, and
  Twamley}}]{said2011nanoscale}
\bibinfo{author}{\bibfnamefont{R.}~\bibnamefont{Said}},
  \bibinfo{author}{\bibfnamefont{D.}~\bibnamefont{Berry}}, \bibnamefont{and}
  \bibinfo{author}{\bibfnamefont{J.}~\bibnamefont{Twamley}},
  \bibinfo{journal}{Physical Review B} \textbf{\bibinfo{volume}{83}},
  \bibinfo{pages}{125410} (\bibinfo{year}{2011}).

\bibitem[{\citenamefont{Higgins et~al.}(2007)\citenamefont{Higgins, Berry,
  Bartlett, Wiseman, and Pryde}}]{higgins2007entanglement}
\bibinfo{author}{\bibfnamefont{B.~L.} \bibnamefont{Higgins}},
  \bibinfo{author}{\bibfnamefont{D.~W.} \bibnamefont{Berry}},
  \bibinfo{author}{\bibfnamefont{S.~D.} \bibnamefont{Bartlett}},
  \bibinfo{author}{\bibfnamefont{H.~M.} \bibnamefont{Wiseman}},
  \bibnamefont{and} \bibinfo{author}{\bibfnamefont{G.~J.} \bibnamefont{Pryde}},
  \bibinfo{journal}{Nature} \textbf{\bibinfo{volume}{450}},
  \bibinfo{pages}{393} (\bibinfo{year}{2007}).

\bibitem[{\citenamefont{Waldherr et~al.}(2012)\citenamefont{Waldherr, Beck,
  Neumann, Said, Nitsche, Markham, Twitchen, Twamley, Jelezko, and
  Wrachtrup}}]{waldherr2012high}
\bibinfo{author}{\bibfnamefont{G.}~\bibnamefont{Waldherr}},
  \bibinfo{author}{\bibfnamefont{J.}~\bibnamefont{Beck}},
  \bibinfo{author}{\bibfnamefont{P.}~\bibnamefont{Neumann}},
  \bibinfo{author}{\bibfnamefont{R.}~\bibnamefont{Said}},
  \bibinfo{author}{\bibfnamefont{M.}~\bibnamefont{Nitsche}},
  \bibinfo{author}{\bibfnamefont{M.}~\bibnamefont{Markham}},
  \bibinfo{author}{\bibfnamefont{D.}~\bibnamefont{Twitchen}},
  \bibinfo{author}{\bibfnamefont{J.}~\bibnamefont{Twamley}},
  \bibinfo{author}{\bibfnamefont{F.}~\bibnamefont{Jelezko}}, \bibnamefont{and}
  \bibinfo{author}{\bibfnamefont{J.}~\bibnamefont{Wrachtrup}},
  \bibinfo{journal}{Nature nanotechnology} \textbf{\bibinfo{volume}{7}},
  \bibinfo{pages}{105} (\bibinfo{year}{2012}).

\bibitem[{\citenamefont{Demkowicz-Dobrzanski
  et~al.}(2012)\citenamefont{Demkowicz-Dobrzanski, Kolodynski, and
  Guta}}]{demkowicz2012elusive}
\bibinfo{author}{\bibfnamefont{R.}~\bibnamefont{Demkowicz-Dobrzanski}},
  \bibinfo{author}{\bibfnamefont{J.}~\bibnamefont{Kolodynski}},
  \bibnamefont{and} \bibinfo{author}{\bibfnamefont{M.}~\bibnamefont{Guta}},
  \bibinfo{journal}{Nature Communications} \textbf{\bibinfo{volume}{3}}
  (\bibinfo{year}{2012}).

\bibitem[{\citenamefont{Gottesman}(2009)}]{gottesman2009introduction}
\bibinfo{author}{\bibfnamefont{D.}~\bibnamefont{Gottesman}},
  \bibinfo{journal}{Proceedings of Symposia in Applied Mathematics}
  \textbf{\bibinfo{volume}{68}}, \bibinfo{pages}{13} (\bibinfo{year}{2009}).

\bibitem[{\citenamefont{Viola et~al.}(1999)\citenamefont{Viola, Knill, and
  Lloyd}}]{viola1999dynamical}
\bibinfo{author}{\bibfnamefont{L.}~\bibnamefont{Viola}},
  \bibinfo{author}{\bibfnamefont{E.}~\bibnamefont{Knill}}, \bibnamefont{and}
  \bibinfo{author}{\bibfnamefont{S.}~\bibnamefont{Lloyd}},
  \bibinfo{journal}{Phys. Rev. Lett.} \textbf{\bibinfo{volume}{82}},
  \bibinfo{pages}{2417} (\bibinfo{year}{1999}).

\bibitem[{\citenamefont{Taylor et~al.}(2008)\citenamefont{Taylor, Cappellaro,
  Childress, Jiang, Budker, Hemmer, Yacoby, Walsworth, and
  Lukin}}]{taylor2008high}
\bibinfo{author}{\bibfnamefont{J.}~\bibnamefont{Taylor}},
  \bibinfo{author}{\bibfnamefont{P.}~\bibnamefont{Cappellaro}},
  \bibinfo{author}{\bibfnamefont{L.}~\bibnamefont{Childress}},
  \bibinfo{author}{\bibfnamefont{L.}~\bibnamefont{Jiang}},
  \bibinfo{author}{\bibfnamefont{D.}~\bibnamefont{Budker}},
  \bibinfo{author}{\bibfnamefont{P.}~\bibnamefont{Hemmer}},
  \bibinfo{author}{\bibfnamefont{A.}~\bibnamefont{Yacoby}},
  \bibinfo{author}{\bibfnamefont{R.}~\bibnamefont{Walsworth}},
  \bibnamefont{and} \bibinfo{author}{\bibfnamefont{M.}~\bibnamefont{Lukin}},
  \bibinfo{journal}{Nature Physics} \textbf{\bibinfo{volume}{4}},
  \bibinfo{pages}{810} (\bibinfo{year}{2008}).

\bibitem[{\citenamefont{De~Lange et~al.}(2011)\citenamefont{De~Lange,
  Rist{\`e}, Dobrovitski, and Hanson}}]{de2011single}
\bibinfo{author}{\bibfnamefont{G.}~\bibnamefont{De~Lange}},
  \bibinfo{author}{\bibfnamefont{D.}~\bibnamefont{Rist{\`e}}},
  \bibinfo{author}{\bibfnamefont{V.}~\bibnamefont{Dobrovitski}},
  \bibnamefont{and} \bibinfo{author}{\bibfnamefont{R.}~\bibnamefont{Hanson}},
  \bibinfo{journal}{Phys. Rev. Lett.} \textbf{\bibinfo{volume}{106}},
  \bibinfo{pages}{080802} (\bibinfo{year}{2011}).

\bibitem[{\citenamefont{D{\"u}r et~al.}(2014)\citenamefont{D{\"u}r,
  Skotiniotis, Froewis, and Kraus}}]{dur2014improved}
\bibinfo{author}{\bibfnamefont{W.}~\bibnamefont{D{\"u}r}},
  \bibinfo{author}{\bibfnamefont{M.}~\bibnamefont{Skotiniotis}},
  \bibinfo{author}{\bibfnamefont{F.}~\bibnamefont{Froewis}}, \bibnamefont{and}
  \bibinfo{author}{\bibfnamefont{B.}~\bibnamefont{Kraus}},
  \bibinfo{journal}{Phys. Rev. Lett.} \textbf{\bibinfo{volume}{112}},
  \bibinfo{pages}{080801} (\bibinfo{year}{2014}).

\bibitem[{\citenamefont{Herrera-Mart{\'\i}
  et~al.}(2015)\citenamefont{Herrera-Mart{\'\i}, Gefen, Aharonov, Katz, and
  Retzker}}]{herrera2015quantum}
\bibinfo{author}{\bibfnamefont{D.~A.} \bibnamefont{Herrera-Mart{\'\i}}},
  \bibinfo{author}{\bibfnamefont{T.}~\bibnamefont{Gefen}},
  \bibinfo{author}{\bibfnamefont{D.}~\bibnamefont{Aharonov}},
  \bibinfo{author}{\bibfnamefont{N.}~\bibnamefont{Katz}}, \bibnamefont{and}
  \bibinfo{author}{\bibfnamefont{A.}~\bibnamefont{Retzker}},
  \bibinfo{journal}{Phys. Rev. Lett.} \textbf{\bibinfo{volume}{115}},
  \bibinfo{pages}{200501} (\bibinfo{year}{2015}).

\bibitem[{\citenamefont{Arrad et~al.}(2014)\citenamefont{Arrad, Vinkler,
  Aharonov, and Retzker}}]{arrad2014increasing}
\bibinfo{author}{\bibfnamefont{G.}~\bibnamefont{Arrad}},
  \bibinfo{author}{\bibfnamefont{Y.}~\bibnamefont{Vinkler}},
  \bibinfo{author}{\bibfnamefont{D.}~\bibnamefont{Aharonov}}, \bibnamefont{and}
  \bibinfo{author}{\bibfnamefont{A.}~\bibnamefont{Retzker}},
  \bibinfo{journal}{Phys. Rev. Lett.} \textbf{\bibinfo{volume}{112}},
  \bibinfo{pages}{150801} (\bibinfo{year}{2014}).

\bibitem[{\citenamefont{Kessler et~al.}(2014)\citenamefont{Kessler, Lovchinsky,
  Sushkov, and Lukin}}]{kessler2014quantum}
\bibinfo{author}{\bibfnamefont{E.~M.} \bibnamefont{Kessler}},
  \bibinfo{author}{\bibfnamefont{I.}~\bibnamefont{Lovchinsky}},
  \bibinfo{author}{\bibfnamefont{A.~O.} \bibnamefont{Sushkov}},
  \bibnamefont{and} \bibinfo{author}{\bibfnamefont{M.~D.} \bibnamefont{Lukin}},
  \bibinfo{journal}{Phys. Rev. Lett.} \textbf{\bibinfo{volume}{112}},
  \bibinfo{pages}{150802} (\bibinfo{year}{2014}).

\bibitem[{\citenamefont{Cohen et~al.}(2016)\citenamefont{Cohen, Pilnyak,
  Istrati, Retzker, and Eisenberg}}]{cohen2016demonstration}
\bibinfo{author}{\bibfnamefont{L.}~\bibnamefont{Cohen}},
  \bibinfo{author}{\bibfnamefont{Y.}~\bibnamefont{Pilnyak}},
  \bibinfo{author}{\bibfnamefont{D.}~\bibnamefont{Istrati}},
  \bibinfo{author}{\bibfnamefont{A.}~\bibnamefont{Retzker}}, \bibnamefont{and}
  \bibinfo{author}{\bibfnamefont{H.}~\bibnamefont{Eisenberg}},
  \bibinfo{journal}{Phys. Rev. A} \textbf{\bibinfo{volume}{94}},
  \bibinfo{pages}{012324} (\bibinfo{year}{2016}).

\bibitem[{\citenamefont{Unden et~al.}(2016)\citenamefont{Unden,
  Balasubramanian, Louzon, Vinkler, Plenio, Markham, Twitchen, Stacey,
  Lovchinsky, Sushkov et~al.}}]{unden2016quantum}
\bibinfo{author}{\bibfnamefont{T.}~\bibnamefont{Unden}},
  \bibinfo{author}{\bibfnamefont{P.}~\bibnamefont{Balasubramanian}},
  \bibinfo{author}{\bibfnamefont{D.}~\bibnamefont{Louzon}},
  \bibinfo{author}{\bibfnamefont{Y.}~\bibnamefont{Vinkler}},
  \bibinfo{author}{\bibfnamefont{M.~B.} \bibnamefont{Plenio}},
  \bibinfo{author}{\bibfnamefont{M.}~\bibnamefont{Markham}},
  \bibinfo{author}{\bibfnamefont{D.}~\bibnamefont{Twitchen}},
  \bibinfo{author}{\bibfnamefont{A.}~\bibnamefont{Stacey}},
  \bibinfo{author}{\bibfnamefont{I.}~\bibnamefont{Lovchinsky}},
  \bibinfo{author}{\bibfnamefont{A.~O.} \bibnamefont{Sushkov}},
  \bibnamefont{et~al.}, \bibinfo{journal}{Phys. Rev. Lett.}
  \textbf{\bibinfo{volume}{116}}, \bibinfo{pages}{230502}
  (\bibinfo{year}{2016}).

\bibitem[{\citenamefont{Sisi and {\it{et al}}}(2017)}]{aqec2017hl}
\bibinfo{author}{\bibfnamefont{Z.}~\bibnamefont{Sisi}} \bibnamefont{and}
  \bibinfo{author}{\bibnamefont{{\it{et al}}}}, \bibinfo{journal}{arXiv
  preprint arXiv:1706.02445}  (\bibinfo{year}{2017}).

\bibitem[{\citenamefont{Schmitt and {\it{et al}}}(2017)}]{asimon2017quantum}
\bibinfo{author}{\bibfnamefont{S.}~\bibnamefont{Schmitt}} \bibnamefont{and}
  \bibinfo{author}{\bibnamefont{{\it{et al}}}}, \bibinfo{journal}{Science}
  \textbf{\bibinfo{volume}{356}}, \bibinfo{pages}{832} (\bibinfo{year}{2017}).

\bibitem[{\citenamefont{Boss et~al.}(2017)\citenamefont{Boss, Cujia, Zopes, and
  Degen}}]{boss2017quantum}
\bibinfo{author}{\bibfnamefont{J.}~\bibnamefont{Boss}},
  \bibinfo{author}{\bibfnamefont{K.}~\bibnamefont{Cujia}},
  \bibinfo{author}{\bibfnamefont{J.}~\bibnamefont{Zopes}}, \bibnamefont{and}
  \bibinfo{author}{\bibfnamefont{C.}~\bibnamefont{Degen}},
  \bibinfo{journal}{Science} \textbf{\bibinfo{volume}{356}},
  \bibinfo{pages}{837} (\bibinfo{year}{2017}).

\bibitem[{\citenamefont{Misra and Sudarshan}(1977)}]{misra1977zeno}
\bibinfo{author}{\bibfnamefont{B.}~\bibnamefont{Misra}} \bibnamefont{and}
  \bibinfo{author}{\bibfnamefont{E.~G.} \bibnamefont{Sudarshan}},
  \bibinfo{journal}{Journal of Mathematical Physics}
  \textbf{\bibinfo{volume}{18}}, \bibinfo{pages}{756} (\bibinfo{year}{1977}).

\bibitem[{\citenamefont{Itano et~al.}(1990)\citenamefont{Itano, Heinzen,
  Bollinger, and Wineland}}]{itano1990quantum}
\bibinfo{author}{\bibfnamefont{W.}~\bibnamefont{Itano}},
  \bibinfo{author}{\bibfnamefont{D.}~\bibnamefont{Heinzen}},
  \bibinfo{author}{\bibfnamefont{J.}~\bibnamefont{Bollinger}},
  \bibnamefont{and} \bibinfo{author}{\bibfnamefont{D.}~\bibnamefont{Wineland}},
  \bibinfo{journal}{Phys. Rev. A} \textbf{\bibinfo{volume}{41}},
  \bibinfo{pages}{2295} (\bibinfo{year}{1990}).

\bibitem[{\citenamefont{Facchi et~al.}(2001)\citenamefont{Facchi, Nakazato, and
  Pascazio}}]{FacchiNakazatoPascazio01a}
\bibinfo{author}{\bibfnamefont{P.}~\bibnamefont{Facchi}},
  \bibinfo{author}{\bibfnamefont{H.}~\bibnamefont{Nakazato}}, \bibnamefont{and}
  \bibinfo{author}{\bibfnamefont{S.}~\bibnamefont{Pascazio}},
  \bibinfo{journal}{Phys. Rev. Lett.} \textbf{\bibinfo{volume}{86}},
  \bibinfo{pages}{2699} (\bibinfo{year}{2001}).

\bibitem[{\citenamefont{Nakazato et~al.}(1996)\citenamefont{Nakazato, Namiki,
  and Pascazio}}]{NakazatoNamikiPascazio01a}
\bibinfo{author}{\bibfnamefont{H.}~\bibnamefont{Nakazato}},
  \bibinfo{author}{\bibfnamefont{M.}~\bibnamefont{Namiki}}, \bibnamefont{and}
  \bibinfo{author}{\bibfnamefont{S.}~\bibnamefont{Pascazio}},
  \bibinfo{journal}{Int. J. Mod. B} \textbf{\bibinfo{volume}{10}},
  \bibinfo{pages}{247} (\bibinfo{year}{1996}).

\bibitem[{\citenamefont{Raussendorf and Briegel}(2001)}]{Raussendorf:2001p368}
\bibinfo{author}{\bibfnamefont{R.}~\bibnamefont{Raussendorf}} \bibnamefont{and}
  \bibinfo{author}{\bibfnamefont{H.}~\bibnamefont{Briegel}},
  \bibinfo{journal}{Phys. Rev. Lett.} \textbf{\bibinfo{volume}{86}},
  \bibinfo{pages}{5188} (\bibinfo{year}{2001}).

\bibitem[{\citenamefont{Barjaktarevic et~al.}(2005)\citenamefont{Barjaktarevic,
  McKenzie, Links, and Milburn}}]{barjaktarevic2005measurement}
\bibinfo{author}{\bibfnamefont{J.}~\bibnamefont{Barjaktarevic}},
  \bibinfo{author}{\bibfnamefont{R.}~\bibnamefont{McKenzie}},
  \bibinfo{author}{\bibfnamefont{J.}~\bibnamefont{Links}}, \bibnamefont{and}
  \bibinfo{author}{\bibfnamefont{G.}~\bibnamefont{Milburn}},
  \bibinfo{journal}{Phys. Rev. Lett.} \textbf{\bibinfo{volume}{95}},
  \bibinfo{pages}{230501} (\bibinfo{year}{2005}).

\bibitem[{\citenamefont{Silva et~al.}(2007)\citenamefont{Silva, Danos, Kashefi,
  and Ollivier}}]{silva2007direct}
\bibinfo{author}{\bibfnamefont{M.}~\bibnamefont{Silva}},
  \bibinfo{author}{\bibfnamefont{V.}~\bibnamefont{Danos}},
  \bibinfo{author}{\bibfnamefont{E.}~\bibnamefont{Kashefi}}, \bibnamefont{and}
  \bibinfo{author}{\bibfnamefont{H.}~\bibnamefont{Ollivier}},
  \bibinfo{journal}{New Journal of Physics} \textbf{\bibinfo{volume}{9}},
  \bibinfo{pages}{192} (\bibinfo{year}{2007}).

\bibitem[{\citenamefont{Olmschenk et~al.}(2009)\citenamefont{Olmschenk,
  Matsukevich, Maunz, Hayes, Duan, and Monroe}}]{olmschenk2009quantum}
\bibinfo{author}{\bibfnamefont{S.}~\bibnamefont{Olmschenk}},
  \bibinfo{author}{\bibfnamefont{D.}~\bibnamefont{Matsukevich}},
  \bibinfo{author}{\bibfnamefont{P.}~\bibnamefont{Maunz}},
  \bibinfo{author}{\bibfnamefont{D.}~\bibnamefont{Hayes}},
  \bibinfo{author}{\bibfnamefont{L.-M.} \bibnamefont{Duan}}, \bibnamefont{and}
  \bibinfo{author}{\bibfnamefont{C.}~\bibnamefont{Monroe}},
  \bibinfo{journal}{Science} \textbf{\bibinfo{volume}{323}},
  \bibinfo{pages}{486} (\bibinfo{year}{2009}).

\bibitem[{\citenamefont{Baur et~al.}(2012)\citenamefont{Baur, Fedorov, Steffen,
  Filipp, Da~Silva, and Wallraff}}]{baur2012benchmarking}
\bibinfo{author}{\bibfnamefont{M.}~\bibnamefont{Baur}},
  \bibinfo{author}{\bibfnamefont{A.}~\bibnamefont{Fedorov}},
  \bibinfo{author}{\bibfnamefont{L.}~\bibnamefont{Steffen}},
  \bibinfo{author}{\bibfnamefont{S.}~\bibnamefont{Filipp}},
  \bibinfo{author}{\bibfnamefont{M.}~\bibnamefont{Da~Silva}}, \bibnamefont{and}
  \bibinfo{author}{\bibfnamefont{A.}~\bibnamefont{Wallraff}},
  \bibinfo{journal}{Phys. Rev. Lett.} \textbf{\bibinfo{volume}{108}},
  \bibinfo{pages}{040502} (\bibinfo{year}{2012}).

\bibitem[{\citenamefont{Averin et~al.}(2016)\citenamefont{Averin, Xu, Zhong,
  Song, Wang, and Han}}]{averin2016suppression}
\bibinfo{author}{\bibfnamefont{D.}~\bibnamefont{Averin}},
  \bibinfo{author}{\bibfnamefont{K.}~\bibnamefont{Xu}},
  \bibinfo{author}{\bibfnamefont{Y.}~\bibnamefont{Zhong}},
  \bibinfo{author}{\bibfnamefont{C.}~\bibnamefont{Song}},
  \bibinfo{author}{\bibfnamefont{H.}~\bibnamefont{Wang}}, \bibnamefont{and}
  \bibinfo{author}{\bibfnamefont{S.}~\bibnamefont{Han}},
  \bibinfo{journal}{Phys. Rev. Lett.} \textbf{\bibinfo{volume}{116}},
  \bibinfo{pages}{010501} (\bibinfo{year}{2016}).

\bibitem[{\citenamefont{Hornberger}(2009)}]{hornberger2009introduction}
\bibinfo{author}{\bibfnamefont{K.}~\bibnamefont{Hornberger}}, in
  \emph{\bibinfo{booktitle}{Entanglement and Decoherence}}
  (\bibinfo{publisher}{Springer}, \bibinfo{year}{2009}), pp.
  \bibinfo{pages}{221--276}.

\bibitem[{\citenamefont{De~Lange et~al.}(2010)\citenamefont{De~Lange, Wang,
  Riste, Dobrovitski, and Hanson}}]{de2010universal}
\bibinfo{author}{\bibfnamefont{G.}~\bibnamefont{De~Lange}},
  \bibinfo{author}{\bibfnamefont{Z.}~\bibnamefont{Wang}},
  \bibinfo{author}{\bibfnamefont{D.}~\bibnamefont{Riste}},
  \bibinfo{author}{\bibfnamefont{V.}~\bibnamefont{Dobrovitski}},
  \bibnamefont{and} \bibinfo{author}{\bibfnamefont{R.}~\bibnamefont{Hanson}},
  \bibinfo{journal}{Science} \textbf{\bibinfo{volume}{330}},
  \bibinfo{pages}{60} (\bibinfo{year}{2010}).

\bibitem[{\citenamefont{Yoshihara et~al.}(2006)\citenamefont{Yoshihara,
  Harrabi, Niskanen, and Nakamura}}]{YoshiharaHarrabiNiskanenNakamura01a}
\bibinfo{author}{\bibfnamefont{F.}~\bibnamefont{Yoshihara}},
  \bibinfo{author}{\bibfnamefont{K.}~\bibnamefont{Harrabi}},
  \bibinfo{author}{\bibfnamefont{A.}~\bibnamefont{Niskanen}}, \bibnamefont{and}
  \bibinfo{author}{\bibfnamefont{Y.}~\bibnamefont{Nakamura}},
  \bibinfo{journal}{Phys. Rev. Lett.} \textbf{\bibinfo{volume}{97}},
  \bibinfo{pages}{167001} (\bibinfo{year}{2006}).

\bibitem[{\citenamefont{Kakuyanagi et~al.}(2007)\citenamefont{Kakuyanagi, Meno,
  Saito, Nakano, Semba, Takayanagi, Deppe, and
  Shnirman}}]{KakuyanagiMenoSaitoNakanoSembaTakayanagiDeppeShnirman01a}
\bibinfo{author}{\bibfnamefont{K.}~\bibnamefont{Kakuyanagi}},
  \bibinfo{author}{\bibfnamefont{T.}~\bibnamefont{Meno}},
  \bibinfo{author}{\bibfnamefont{S.}~\bibnamefont{Saito}},
  \bibinfo{author}{\bibfnamefont{H.}~\bibnamefont{Nakano}},
  \bibinfo{author}{\bibfnamefont{K.}~\bibnamefont{Semba}},
  \bibinfo{author}{\bibfnamefont{H.}~\bibnamefont{Takayanagi}},
  \bibinfo{author}{\bibfnamefont{F.}~\bibnamefont{Deppe}}, \bibnamefont{and}
  \bibinfo{author}{\bibfnamefont{A.}~\bibnamefont{Shnirman}},
  \bibinfo{journal}{Phys. Rev. Lett.} \textbf{\bibinfo{volume}{98}},
  \bibinfo{pages}{047004} (\bibinfo{year}{2007}).

\bibitem[{\citenamefont{Kondo et~al.}(2016)\citenamefont{Kondo, Matsuzaki,
  Matsushima, and Filgueiras}}]{kondo2016using}
\bibinfo{author}{\bibfnamefont{Y.}~\bibnamefont{Kondo}},
  \bibinfo{author}{\bibfnamefont{Y.}~\bibnamefont{Matsuzaki}},
  \bibinfo{author}{\bibfnamefont{K.}~\bibnamefont{Matsushima}},
  \bibnamefont{and} \bibinfo{author}{\bibfnamefont{J.~G.}
  \bibnamefont{Filgueiras}}, \bibinfo{journal}{New Journal of Physics}
  \textbf{\bibinfo{volume}{18}}, \bibinfo{pages}{013033}
  (\bibinfo{year}{2016}).

\bibitem[{\citenamefont{Palma et~al.}(1996)\citenamefont{Palma, Suominen, and
  Ekert}}]{PSE}
\bibinfo{author}{\bibfnamefont{G.~M.} \bibnamefont{Palma}},
  \bibinfo{author}{\bibfnamefont{K.~A.} \bibnamefont{Suominen}},
  \bibnamefont{and} \bibinfo{author}{\bibfnamefont{A.~K.} \bibnamefont{Ekert}},
  \bibinfo{journal}{Proc. R. Soc. London. Ser.A}
  \textbf{\bibinfo{volume}{452}}, \bibinfo{pages}{567} (\bibinfo{year}{1996}).

\bibitem[{\citenamefont{Browne and Briegel}(2006)}]{browne-2006-}
\bibinfo{author}{\bibfnamefont{D.~E.} \bibnamefont{Browne}} \bibnamefont{and}
  \bibinfo{author}{\bibfnamefont{H.~J.} \bibnamefont{Briegel}}
  (\bibinfo{year}{2006}), \bibinfo{note}{quant-ph/0603226}.

\bibitem[{\citenamefont{Schaffry et~al.}(2010)\citenamefont{Schaffry, Gauger,
  Morton, Fitzsimons, Benjamin, and Lovett}}]{schaffry2010quantum}
\bibinfo{author}{\bibfnamefont{M.}~\bibnamefont{Schaffry}},
  \bibinfo{author}{\bibfnamefont{E.~M.} \bibnamefont{Gauger}},
  \bibinfo{author}{\bibfnamefont{J.~J.} \bibnamefont{Morton}},
  \bibinfo{author}{\bibfnamefont{J.}~\bibnamefont{Fitzsimons}},
  \bibinfo{author}{\bibfnamefont{S.~C.} \bibnamefont{Benjamin}},
  \bibnamefont{and} \bibinfo{author}{\bibfnamefont{B.~W.}
  \bibnamefont{Lovett}}, \bibinfo{journal}{Physical Review A}
  \textbf{\bibinfo{volume}{82}}, \bibinfo{pages}{042114}
  (\bibinfo{year}{2010}).

\bibitem[{\citenamefont{Jones et~al.}(2009)\citenamefont{Jones, Karlen,
  Fitzsimons, Ardavan, Benjamin, Briggs, and Morton}}]{jones2009magnetic}
\bibinfo{author}{\bibfnamefont{J.~A.} \bibnamefont{Jones}},
  \bibinfo{author}{\bibfnamefont{S.~D.} \bibnamefont{Karlen}},
  \bibinfo{author}{\bibfnamefont{J.}~\bibnamefont{Fitzsimons}},
  \bibinfo{author}{\bibfnamefont{A.}~\bibnamefont{Ardavan}},
  \bibinfo{author}{\bibfnamefont{S.~C.} \bibnamefont{Benjamin}},
  \bibinfo{author}{\bibfnamefont{G.~A.~D.} \bibnamefont{Briggs}},
  \bibnamefont{and} \bibinfo{author}{\bibfnamefont{J.~J.}
  \bibnamefont{Morton}}, \bibinfo{journal}{Science}
  \textbf{\bibinfo{volume}{324}}, \bibinfo{pages}{1166} (\bibinfo{year}{2009}).

\bibitem[{\citenamefont{Matsuzaki et~al.}(2011)\citenamefont{Matsuzaki,
  Benjamin, and Fitzsimons}}]{matsuzaki2011magnetic}
\bibinfo{author}{\bibfnamefont{Y.}~\bibnamefont{Matsuzaki}},
  \bibinfo{author}{\bibfnamefont{S.}~\bibnamefont{Benjamin}}, \bibnamefont{and}
  \bibinfo{author}{\bibfnamefont{J.}~\bibnamefont{Fitzsimons}},
  \bibinfo{journal}{Phys. Rev. A} \textbf{\bibinfo{volume}{84}},
  \bibinfo{pages}{012103} (\bibinfo{year}{2011}).

\bibitem[{\citenamefont{Chin et~al.}(2012)\citenamefont{Chin, Huelga, and
  Plenio}}]{chin2012quantum}
\bibinfo{author}{\bibfnamefont{A.~W.} \bibnamefont{Chin}},
  \bibinfo{author}{\bibfnamefont{S.~F.} \bibnamefont{Huelga}},
  \bibnamefont{and} \bibinfo{author}{\bibfnamefont{M.~B.}
  \bibnamefont{Plenio}}, \bibinfo{journal}{Phys. Rev. Lett.}
  \textbf{\bibinfo{volume}{109}}, \bibinfo{pages}{233601}
  (\bibinfo{year}{2012}).

\bibitem[{\citenamefont{Shaji and Caves}(2007)}]{shaji2007qubit}
\bibinfo{author}{\bibfnamefont{A.}~\bibnamefont{Shaji}} \bibnamefont{and}
  \bibinfo{author}{\bibfnamefont{C.~M.} \bibnamefont{Caves}},
  \bibinfo{journal}{Physical Review A} \textbf{\bibinfo{volume}{76}},
  \bibinfo{pages}{032111} (\bibinfo{year}{2007}).

\bibitem[{\citenamefont{Bohmann et~al.}(2015)\citenamefont{Bohmann, Sperling,
  and Vogel}}]{bohmann2015entanglement}
\bibinfo{author}{\bibfnamefont{M.}~\bibnamefont{Bohmann}},
  \bibinfo{author}{\bibfnamefont{J.}~\bibnamefont{Sperling}}, \bibnamefont{and}
  \bibinfo{author}{\bibfnamefont{W.}~\bibnamefont{Vogel}},
  \bibinfo{journal}{Physical Review A} \textbf{\bibinfo{volume}{91}},
  \bibinfo{pages}{042332} (\bibinfo{year}{2015}).

              \bibitem[{\citenamefont{Hornberger}(2009)}]{hornberger2009introduction}
\bibinfo{author}{\bibfnamefont{K.}~\bibnamefont{Hornberger}}, in
  \emph{\bibinfo{booktitle}{Entanglement and Decoherence}}
  (\bibinfo{publisher}{Springer}, \bibinfo{year}{2009}), pp.
  \bibinfo{pages}{221--276}.

\bibitem[{\citenamefont{Hasegawa}(2011)}]{hasegawa2011classical}
\bibinfo{author}{\bibfnamefont{H.}~\bibnamefont{Hasegawa}},
  \bibinfo{journal}{Physical Review E} \textbf{\bibinfo{volume}{83}},
  \bibinfo{pages}{021104} (\bibinfo{year}{2011}).

\bibitem[{\citenamefont{Carcaterra and Akay}(2011)}]{carcaterra2011dissipation}
\bibinfo{author}{\bibfnamefont{A.}~\bibnamefont{Carcaterra}} \bibnamefont{and}
  \bibinfo{author}{\bibfnamefont{A.}~\bibnamefont{Akay}},
  \bibinfo{journal}{Physical Review E} \textbf{\bibinfo{volume}{84}},
  \bibinfo{pages}{011121} (\bibinfo{year}{2011}).

\bibitem[{\citenamefont{Romach et~al.}(2015)\citenamefont{Romach, M{\"u}ller,
  Unden, Rogers, Isoda, Itoh, Markham, Stacey, Meijer, Pezzagna
  et~al.}}]{romach2015spectroscopy}
\bibinfo{author}{\bibfnamefont{Y.}~\bibnamefont{Romach}},
  \bibinfo{author}{\bibfnamefont{C.}~\bibnamefont{M{\"u}ller}},
  \bibinfo{author}{\bibfnamefont{T.}~\bibnamefont{Unden}},
  \bibinfo{author}{\bibfnamefont{L.}~\bibnamefont{Rogers}},
  \bibinfo{author}{\bibfnamefont{T.}~\bibnamefont{Isoda}},
  \bibinfo{author}{\bibfnamefont{K.}~\bibnamefont{Itoh}},
  \bibinfo{author}{\bibfnamefont{M.}~\bibnamefont{Markham}},
  \bibinfo{author}{\bibfnamefont{A.}~\bibnamefont{Stacey}},
  \bibinfo{author}{\bibfnamefont{J.}~\bibnamefont{Meijer}},
  \bibinfo{author}{\bibfnamefont{S.}~\bibnamefont{Pezzagna}},
  \bibnamefont{et~al.}, \bibinfo{journal}{Phys. Rev. Lett.}
  \textbf{\bibinfo{volume}{114}}, \bibinfo{pages}{017601}
  (\bibinfo{year}{2015}).

           \bibitem[{\citenamefont{Palma et~al.}(1996)\citenamefont{Palma, Suominen, and
  Ekert}}]{PSE}
\bibinfo{author}{\bibfnamefont{G.~M.} \bibnamefont{Palma}},
  \bibinfo{author}{\bibfnamefont{K.~A.} \bibnamefont{Suominen}},
  \bibnamefont{and} \bibinfo{author}{\bibfnamefont{A.~K.} \bibnamefont{Ekert}},
  \bibinfo{journal}{Proc. R. Soc. London. Ser.A}
  \textbf{\bibinfo{volume}{452}}, \bibinfo{pages}{567} (\bibinfo{year}{1996}).

\bibitem[{\citenamefont{De~Lange et~al.}(2010)\citenamefont{De~Lange, Wang,
  Riste, Dobrovitski, and Hanson}}]{de2010universal}
\bibinfo{author}{\bibfnamefont{G.}~\bibnamefont{De~Lange}},
  \bibinfo{author}{\bibfnamefont{Z.}~\bibnamefont{Wang}},
  \bibinfo{author}{\bibfnamefont{D.}~\bibnamefont{Riste}},
  \bibinfo{author}{\bibfnamefont{V.}~\bibnamefont{Dobrovitski}},
  \bibnamefont{and} \bibinfo{author}{\bibfnamefont{R.}~\bibnamefont{Hanson}},
  \bibinfo{journal}{Science} \textbf{\bibinfo{volume}{330}},
  \bibinfo{pages}{60} (\bibinfo{year}{2010}).

                 \bibitem[{\citenamefont{Kakuyanagi et~al.}(2007)\citenamefont{Kakuyanagi, Meno,
  Saito, Nakano, Semba, Takayanagi, Deppe, and
  Shnirman}}]{KakuyanagiMenoSaitoNakanoSembaTakayanagiDeppeShnirman01a}
\bibinfo{author}{\bibfnamefont{K.}~\bibnamefont{Kakuyanagi}},
  \bibinfo{author}{\bibfnamefont{T.}~\bibnamefont{Meno}},
  \bibinfo{author}{\bibfnamefont{S.}~\bibnamefont{Saito}},
  \bibinfo{author}{\bibfnamefont{H.}~\bibnamefont{Nakano}},
  \bibinfo{author}{\bibfnamefont{K.}~\bibnamefont{Semba}},
  \bibinfo{author}{\bibfnamefont{H.}~\bibnamefont{Takayanagi}},
  \bibinfo{author}{\bibfnamefont{F.}~\bibnamefont{Deppe}}, \bibnamefont{and}
  \bibinfo{author}{\bibfnamefont{A.}~\bibnamefont{Shnirman}},
  \bibinfo{journal}{Phys. Rev. Lett.} \textbf{\bibinfo{volume}{98}},
  \bibinfo{pages}{047004} (\bibinfo{year}{2007}).

             \bibitem[{\citenamefont{Yoshihara et~al.}(2006)\citenamefont{Yoshihara,
  Harrabi, Niskanen, and Nakamura}}]{YoshiharaHarrabiNiskanenNakamura01a}
\bibinfo{author}{\bibfnamefont{F.}~\bibnamefont{Yoshihara}},
  \bibinfo{author}{\bibfnamefont{K.}~\bibnamefont{Harrabi}},
  \bibinfo{author}{\bibfnamefont{A.}~\bibnamefont{Niskanen}}, \bibnamefont{and}
  \bibinfo{author}{\bibfnamefont{Y.}~\bibnamefont{Nakamura}},
  \bibinfo{journal}{Phys. Rev. Lett.} \textbf{\bibinfo{volume}{97}},
  \bibinfo{pages}{167001} (\bibinfo{year}{2006}).

\bibitem[{\citenamefont{Kondo et~al.}(2016)\citenamefont{Kondo, Matsuzaki,
  Matsushima, and Filgueiras}}]{kondo2016using}
\bibinfo{author}{\bibfnamefont{Y.}~\bibnamefont{Kondo}},
  \bibinfo{author}{\bibfnamefont{Y.}~\bibnamefont{Matsuzaki}},
  \bibinfo{author}{\bibfnamefont{K.}~\bibnamefont{Matsushima}},
  \bibnamefont{and} \bibinfo{author}{\bibfnamefont{J.~G.}
  \bibnamefont{Filgueiras}}, \bibinfo{journal}{New Journal of Physics}
  \textbf{\bibinfo{volume}{18}}, \bibinfo{pages}{013033}
  (\bibinfo{year}{2016}).

\bibitem[{\citenamefont{Wilhelm et~al.}(2007)\citenamefont{Wilhelm, Storcz,
  Hartmann, and Geller}}]{wilhelm2007superconducting}
\bibinfo{author}{\bibfnamefont{F.}~\bibnamefont{Wilhelm}},
  \bibinfo{author}{\bibfnamefont{M.}~\bibnamefont{Storcz}},
  \bibinfo{author}{\bibfnamefont{U.}~\bibnamefont{Hartmann}}, \bibnamefont{and}
  \bibinfo{author}{\bibfnamefont{M.~R.} \bibnamefont{Geller}}, in
  \emph{\bibinfo{booktitle}{Manipulating Quantum Coherence in Solid State
  Systems}} (\bibinfo{publisher}{Springer}, \bibinfo{year}{2007}), pp.
  \bibinfo{pages}{195--232}.

\bibitem[{\citenamefont{Hall et~al.}(2014)\citenamefont{Hall, Cole, and
  Hollenberg}}]{hall2014analytic}
\bibinfo{author}{\bibfnamefont{L.~T.} \bibnamefont{Hall}},
  \bibinfo{author}{\bibfnamefont{J.~H.} \bibnamefont{Cole}}, \bibnamefont{and}
  \bibinfo{author}{\bibfnamefont{L.~C.} \bibnamefont{Hollenberg}},
  \bibinfo{journal}{Physical Review B} \textbf{\bibinfo{volume}{90}},
  \bibinfo{pages}{075201} (\bibinfo{year}{2014}).

\bibitem[{\citenamefont{Koppens et~al.}(2007)\citenamefont{Koppens, Klauser,
  Coish, Nowack, Kouwenhoven, Loss, and Vandersypen}}]{koppens2007universal}
\bibinfo{author}{\bibfnamefont{F.}~\bibnamefont{Koppens}},
  \bibinfo{author}{\bibfnamefont{D.}~\bibnamefont{Klauser}},
  \bibinfo{author}{\bibfnamefont{W.}~\bibnamefont{Coish}},
  \bibinfo{author}{\bibfnamefont{K.}~\bibnamefont{Nowack}},
  \bibinfo{author}{\bibfnamefont{L.}~\bibnamefont{Kouwenhoven}},
  \bibinfo{author}{\bibfnamefont{D.}~\bibnamefont{Loss}}, \bibnamefont{and}
  \bibinfo{author}{\bibfnamefont{L.}~\bibnamefont{Vandersypen}},
  \bibinfo{journal}{Phys. Rev. Lett.} \textbf{\bibinfo{volume}{99}},
  \bibinfo{pages}{106803} (\bibinfo{year}{2007}).

\bibitem[{\citenamefont{Gardiner and Zoller}(2004)}]{GZ01b}
\bibinfo{author}{\bibfnamefont{C.~W.} \bibnamefont{Gardiner}} \bibnamefont{and}
  \bibinfo{author}{\bibfnamefont{P.}~\bibnamefont{Zoller}},
  \emph{\bibinfo{title}{Quantum Noise}} (\bibinfo{publisher}{Springer, Berlin},
  \bibinfo{year}{2004}).

\bibitem[{\citenamefont{Barends et~al.}(2014)\citenamefont{Barends, Kelly,
  Megrant, Veitia, Sank, Jeffrey, White, Mutus, Fowler, Campbell
  et~al.}}]{barends2014superconducting}
\bibinfo{author}{\bibfnamefont{R.}~\bibnamefont{Barends}},
  \bibinfo{author}{\bibfnamefont{J.}~\bibnamefont{Kelly}},
  \bibinfo{author}{\bibfnamefont{A.}~\bibnamefont{Megrant}},
  \bibinfo{author}{\bibfnamefont{A.}~\bibnamefont{Veitia}},
  \bibinfo{author}{\bibfnamefont{D.}~\bibnamefont{Sank}},
  \bibinfo{author}{\bibfnamefont{E.}~\bibnamefont{Jeffrey}},
  \bibinfo{author}{\bibfnamefont{T.~C.} \bibnamefont{White}},
  \bibinfo{author}{\bibfnamefont{J.}~\bibnamefont{Mutus}},
  \bibinfo{author}{\bibfnamefont{A.~G.} \bibnamefont{Fowler}},
  \bibinfo{author}{\bibfnamefont{B.}~\bibnamefont{Campbell}},
  \bibnamefont{et~al.}, \bibinfo{journal}{Nature}
  \textbf{\bibinfo{volume}{508}}, \bibinfo{pages}{500} (\bibinfo{year}{2014}).

\bibitem[{\citenamefont{Ballance et~al.}(2016)\citenamefont{Ballance, Harty,
  Linke, Sepiol, and Lucas}}]{ballance2016high}
\bibinfo{author}{\bibfnamefont{C.}~\bibnamefont{Ballance}},
  \bibinfo{author}{\bibfnamefont{T.}~\bibnamefont{Harty}},
  \bibinfo{author}{\bibfnamefont{N.}~\bibnamefont{Linke}},
  \bibinfo{author}{\bibfnamefont{M.}~\bibnamefont{Sepiol}}, \bibnamefont{and}
  \bibinfo{author}{\bibfnamefont{D.}~\bibnamefont{Lucas}},
  \bibinfo{journal}{Phys. Rev. Lett.} \textbf{\bibinfo{volume}{117}},
  \bibinfo{pages}{060504} (\bibinfo{year}{2016}).

\bibitem[{\citenamefont{Blume-Kohout et~al.}(2017)\citenamefont{Blume-Kohout,
  Gamble, Nielsen, Rudinger, Mizrahi, Fortier, and
  Maunz}}]{blume2017demonstration}
\bibinfo{author}{\bibfnamefont{R.}~\bibnamefont{Blume-Kohout}},
  \bibinfo{author}{\bibfnamefont{J.~K.} \bibnamefont{Gamble}},
  \bibinfo{author}{\bibfnamefont{E.}~\bibnamefont{Nielsen}},
  \bibinfo{author}{\bibfnamefont{K.}~\bibnamefont{Rudinger}},
  \bibinfo{author}{\bibfnamefont{J.}~\bibnamefont{Mizrahi}},
  \bibinfo{author}{\bibfnamefont{K.}~\bibnamefont{Fortier}}, \bibnamefont{and}
  \bibinfo{author}{\bibfnamefont{P.}~\bibnamefont{Maunz}},
  \bibinfo{journal}{Nature Communications} \textbf{\bibinfo{volume}{8}}
  (\bibinfo{year}{2017}).

\bibitem[{\citenamefont{Neill et~al.}(2017)\citenamefont{Neill, Roushan,
  Kechedzhi, Boixo, Isakov, Smelyanskiy, Barends, Burkett, Chen, Chen
  et~al.}}]{neill2017blueprint}
\bibinfo{author}{\bibfnamefont{C.}~\bibnamefont{Neill}},
  \bibinfo{author}{\bibfnamefont{P.}~\bibnamefont{Roushan}},
  \bibinfo{author}{\bibfnamefont{K.}~\bibnamefont{Kechedzhi}},
  \bibinfo{author}{\bibfnamefont{S.}~\bibnamefont{Boixo}},
  \bibinfo{author}{\bibfnamefont{S.}~\bibnamefont{Isakov}},
  \bibinfo{author}{\bibfnamefont{V.}~\bibnamefont{Smelyanskiy}},
  \bibinfo{author}{\bibfnamefont{R.}~\bibnamefont{Barends}},
  \bibinfo{author}{\bibfnamefont{B.}~\bibnamefont{Burkett}},
  \bibinfo{author}{\bibfnamefont{Y.}~\bibnamefont{Chen}},
  \bibinfo{author}{\bibfnamefont{Z.}~\bibnamefont{Chen}}, \bibnamefont{et~al.},
  \bibinfo{journal}{arXiv preprint arXiv:1709.06678}  (\bibinfo{year}{2017}).

\bibitem[{\citenamefont{Popkin}(2016)}]{popkin2016quest}
\bibinfo{author}{\bibfnamefont{G.}~\bibnamefont{Popkin}},
  \bibinfo{journal}{Science} \textbf{\bibinfo{volume}{354}},
  \bibinfo{pages}{1090} (\bibinfo{year}{2016}).

\end{thebibliography}

\end{document}